\documentclass[11pt,a4paper]{article}
\usepackage{a4}
\usepackage{graphicx}
\usepackage{amsfonts}
\usepackage{amssymb}
\usepackage{amsmath}
\usepackage{amssymb}
\usepackage{graphicx}
\usepackage{enumerate}
\usepackage{stmaryrd} 
\usepackage{array}
\usepackage{delarray}
\usepackage{epsfig}
\usepackage{bbm}
\usepackage{color} 
\usepackage{psfrag}
\usepackage{amsmath,amsthm}

\def\div{{\rm div }}

\def\R{\mathbb{R}}

\def\E{\mathbb{E}}

\def\P{\mathbb{P}}

\newtheorem{proposition}{Proposition}

\DeclareMathOperator*{\argmin}{{\rm argmin}}

\newcommand{\un}{{\mathbbm{1}}}

\title{A multiple replica approach to simulate reactive trajectories}
\author{Fr\'ed\'eric C\'erou\thanks{INRIA Rennes - Bretagne Atlantique, Campus de Beaulieu, 35042 Rennes Cedex, France. Frederic.Cerou@inria.fr}, Arnaud Guyader\thanks{INRIA Rennes - Bretagne Atlantique and Universit\'e de Haute Bretagne,
Place du Recteur H. Le Moal, CS 24307,
35043 Rennes Cedex, France. arnaud.guyader@uhb.fr}, Tony Leli\`evre\thanks{CERMICS, Ecole des Ponts ParisTech, 6-8 avenue Blaise Pascal, 77455 Marne La Vall\'ee, France. tony.lelievre@cermics.encp.fr}, David Pommier\thanks{CERMICS, Ecole des Ponts ParisTech, 6-8 avenue Blaise Pascal, 77455 Marne La Vall\'ee, France. david.pommier@cermics.encp.fr}}
\begin{document}
\maketitle
\begin{abstract}
A method to generate reactive trajectories, namely equilibrium trajectories leaving a metastable state and ending in another one is proposed. The algorithm is based on simulating in parallel many copies of the system, and selecting the replicas which have reached the highest values along a chosen one-dimensional reaction coordinate. This reaction coordinate does not need to precisely describe all the metastabilities of the system for the method to give reliable results. An extension of the algorithm to compute transition times from one metastable state to another one is also presented.

We demonstrate the interest of the method on two simple cases: a one-dimensional two-well potential and a two-dimensional potential exhibiting two channels to pass from one metastable state to another one.
\end{abstract}

\section{Introduction}
\label{sec:intro}

A very challenging problem in molecular dynamics is to compute reactive paths, namely trajectories of the system leaving a given metastable state (say $A$, an open subset of $\R^d$), and ending in another one (say $B$, another open subset of $\R^d$ such that $A \cap B = \emptyset$), without going back to $A$. The difficulty comes from the fact that a dynamics at equilibrium typically remains for a very long time around a metastable state before hoping to another one. In other words, most of the trajectories leaving $A$ will go back to $A$, rather than reaching $B$. There exist many methods to sample the canonical equilibrium measure in such a situation, and compute equilibrium thermodynamics quantities like for example the free energy (see~\cite{chipot-pohorille-07,lelievre-rousset-stoltz-10} and references therein) but it is much more difficult to compute dynamical quantities at equilibrium along reactive paths, like transport coefficients and transition rates.

In such situations, it is very often the case that a reaction coordinate is known, which in some sense, indexes transitions from $A$ to $B$. In this paper, a reaction coordinate is meant to be a smooth one-dimensional function:
$$\xi : \R^d \to \R$$
such that:
\begin{equation}\label{eq:hyp_RC}
|\nabla \xi| \neq 0,~ A \subset \{x \in \R^d, \xi(x) < z_{\rm min} \} \text{ and } B \subset \{x \in \R^d, \xi(x) > z_{\rm max} \},
\end{equation}
where $z_{\rm min}<z_{\rm max}$ are two given real numbers. Let us denote $$\Sigma_z=\{ x \in \R^d, \xi(x)=z\}$$ the submanifold of configurations at a fixed value $z$ of the reaction coordinate. For the algorithm we propose to give reliable results, one needs $\Sigma_{z_{\rm min}}$ (resp. $\Sigma_{z_{\rm max}}$) to be ``sufficiently close'' to $A$ (resp. $B$). More precisely, we require that most trajectories starting from the submanifold $\Sigma_{z_{\rm min}}$ (resp. $\Sigma_{z_{\rm max}}$) ends in $A$ (resp. in $B$).

The algorithm we propose in this paper is inspired from methods used in statistics to deal with rare events simulations and estimations. This approach, known as {\em multilevel splitting}, dates back to Kahn and Harris~\cite{kahn-harris-51} and Rosenbluth and Rosenbluth~\cite{rosenbluth-rosenbluth-55}. All the variants share the same basic idea, that is discarding the failed trajectories, and splitting (or branching) the trajectories approaching the rare set. We refer the reader to~\cite{glasserman-heidelberger-shahabuddin-zajic-99} for a review of the multilevel splitting method and a list of references. We will mainly focus here on adaptive variants of this method which have been recently proposed in this domain~\cite{cerou-guyader-07,guyader-hengartner-matzner-10}.

In our context, the idea is to perform an iterative process on many replicas of trajectories which start from the metastable region $A$, and end either in $A$ or in~$B$, and to kill progressively the trajectories which have not reached high values along $\xi$. At the end, an equilibrium ensemble of trajectories starting from $A$ and ending in $B$ are obtained, with a bias which scales like $O(1/N)$, $N$ being the number of trajectories. This scaling of the bias is classical for Monte Carlo methods based on interacting replicas~\cite{del-moral-04}, and is typically negligible compared to the statistical noise which scales like $O(1/\sqrt{N})$. Compared to a brute force algorithm, the computational cost is typically reduced by a factor 1\,000 (see Section~\ref{sec:num} for more details). The details of the algorithm are provided in Section~\ref{sec:algo}.

One of the differences between the algorithm we propose and the transition interface sampling method~\cite{van-erp-moroni-bolhuis-03,van-erp-bolhuis-05},  the forward flux sampling method~\cite{allen-warren-ten-wolde-05,allen-valeriani-ten-wolde-09}, or the milestoning method~\cite{faradjian-elber-04,maragliano-vanden-eijnden-roux-09} whose aim is also to compute reactive trajectories through paths ensembles, is that we do not need to decide {\em a priori} of a given discrete set of values $z_{\rm min}=z_0 < z_1 < z_2 < \ldots < z_n = z_{\rm max}$ through which the trajectories (more precisely, the reaction coordinate along the trajectories) will go. In some sense, these are {\em adaptively} selected by the algorithm, with typically fine discretizations in regions with high gradients of the potential energy, before saddle points, and coarser discretizations in flat regions. Other techniques to sample reactive trajectories include the string method~\cite{e-vanden-eijnden-04,e-ren-vanden-eijnden-02,e-ren-vanden-eijnden-05}, see also the review paper~\cite{dellago-bolhuis-09}.

The main interests of the algorithm we propose are: (i) It does not require  fine tuning of any numerical parameter, nor {\em a priori} discretization of the reaction coordinate values; (ii) It can be applied to any Markovian stochastic dynamics (overdamped Langevin, Langevin, Hybrid Monte Carlo, etc.); (iii) It is reliable whatever the chosen reaction coordinate satisfying~\eqref{eq:hyp_RC}, and in particular if the reaction coordinate does not describe all the metastabilities. This is for example the case if, conditionally to a given value of $\xi$, the canonical measure is multimodal (or, equivalently, the potential energy exhibits wells separated by high barriers along some submanifolds $\Sigma_z$). This is actually a generic situation in practice, encountered in particular when multiple pathways link the two metastable states $A$ and $B$ (see Section~\ref{sec:2d} for a numerical illustration).

The article is organized as follows. Section~\ref{sec:algo} describes the algorithm to compute reactive paths. Section~\ref{sec:num} illustrates
the efficiency of the algorithm on two prototypical cases. Section~\ref{sec:4}
proposes an extension of the approach to evaluate mean transition
times, which is illustrated in Section~\ref{sec:nd_T} by some numerical applications. Finally, conclusions and possible extensions are discussed in Section~\ref{sec:conc}.

\section{Computing reactive trajectories: the algorithm}
\label{sec:algo}

\subsection{Reactive trajectories}
Let $V: \R^d \to \R$ denote the potential function, and let us consider, to fix ideas, overdamped Langevin dynamics:
\begin{equation}
\label{eq:sde}
dX_t = - \nabla V (X_t) \, dt + \sqrt{2 \beta^{-1}} dW_t,
\end{equation}
where $\beta=1 / (k_B T)$. As mentioned above and as will become clear below, the algorithm actually applies to any (continuous in time) Markovian stochastic dynamics, like Langevin dynamics, or Hybrid Monte Carlo~\cite{DKPR87} for example.

The equilibrium canonical measure is
$$d\mu = Z^{-1} \exp ( - \beta V(x)) \, dx$$
where $\displaystyle{Z= \int_{\R^d}\exp ( - \beta V(x)) \, dx}$. The
equilibrium trajectories are those obtained with initial conditions $X_0$
distributed according to $\mu$, and which satisfy~\eqref{eq:sde}. We assume
that $A$ and $B$ are two metastable regions for the dynamics~\eqref{eq:sde},
namely a trajectory starting from $A$ (resp. $B$) remains for a long time in a
neighborhood of $A$ (resp. $B$). A reactive trajectory (from $A$ to
$B$)~\cite{hummer-04,metzner-schuette-vanden-eijnden-06} is a portion of an
equilibrium trajectory which links $A$ to $B$. Thus, it is a trajectory which
leaves $A$, does not come back to $A$, and reaches $B$. It is in general difficult to generate such trajectories, and we propose an algorithm to build an ensemble of $N$ reactive trajectories, using a reaction coordinate $\xi$ which satisfies~\eqref{eq:hyp_RC}. In the numerical experiments below, we will test various reaction coordinates, but, to fix ideas, one could think of $\xi(x)=\|x-x_A\|$ where $x_A \in A$ denotes a reference configuration in $A$, and $\|\cdot\|$ is the Euclidean norm.

\subsection{Details of the algorithm}\label{sec:algo_details}

The algorithm starts with an initialization procedure which consists in three steps (see the end of Section~\ref{sec:trans_time_gen} for an alternative equivalent initialization method):
\begin{enumerate}
\item Generate an ensemble of initial conditions $(X_0^n)_{1 \le n \le N}$ distributed according to the canonical measure $\mu$ conditionally to being in $A$, namely:
$$d\mu_A=Z_A^{-1} \exp(-\beta V(x)) \un_A(x) \, dx$$
where $Z_A=\int_A \exp(-\beta V(x)) \, dx$ and $\un_A$ denotes the characteristic function of the region $A$. For this first step, we use a subsampling of a long trajectory satisfying~\eqref{eq:sde}, starting in $A$ and remaining in $A$ through a Metropolis-Hastings procedure.
\item Starting from the initial conditions $(X_0^n)_{1 \le n \le N}$, for each $n \in \{1, \ldots, N\}$, compute the trajectory $(X_t^n)_{t \ge 0}$ satisfying~\eqref{eq:sde} driven by a Brownian motion $W^n_t$, until the stopping time
$$\sigma^n= \inf \{t \ge 0, \xi(X^n_t) \ge z_{\rm min} \}.$$
The Brownian motions $(W^n_t)$ are of course assumed to be independent. Only the end of the trajectory $(X^n_t)_{0 \le t \le \sigma^n}$, between $A$ and $\Sigma_{z_{\rm min}}$, is retained. With a small abuse of notation, let us denote $0$ the last time at which $X^n_t$ leaves $A$ (and $\sigma^n$ accordingly), so that $X^n_0 \in \partial A$, $X^n_{\sigma^n} \in \Sigma_{z_{\rm min}}$ and $X^n_t$ does not touch either $\partial A$ nor $\Sigma_{z_{\rm min}}$ for $ t \in (0,\sigma^n)$. This second step (which can be seen as one iteration of a transition interface sampling procedure) is not computationally demanding since $\Sigma_{z_{\rm min}}$ is assumed to be ``close to'' $A$ so that $\sigma^n$ is typically small.
\item Continue the simulation of $(X^n_t)_{t \ge \sigma^n}$ (according to~\eqref{eq:sde} with $W_t=W^n_t$) until the stopping time
$$\tau^n= \inf \{t \ge\sigma^n, X^n_t \in A \cup B \}.$$
\end{enumerate}

\begin{figure}[htbp]
\psfrag{A}{$A$}
\psfrag{B}{$B$}
\psfrag{Sig1}{$\Sigma_{z_{\rm min}}$}
\psfrag{Sig2}{$\Sigma_{z_{\rm max}}$}
\psfrag{q1}{$q_1$}
\psfrag{q2}{$q_2$}
  \centering
\includegraphics[width=6.5cm]{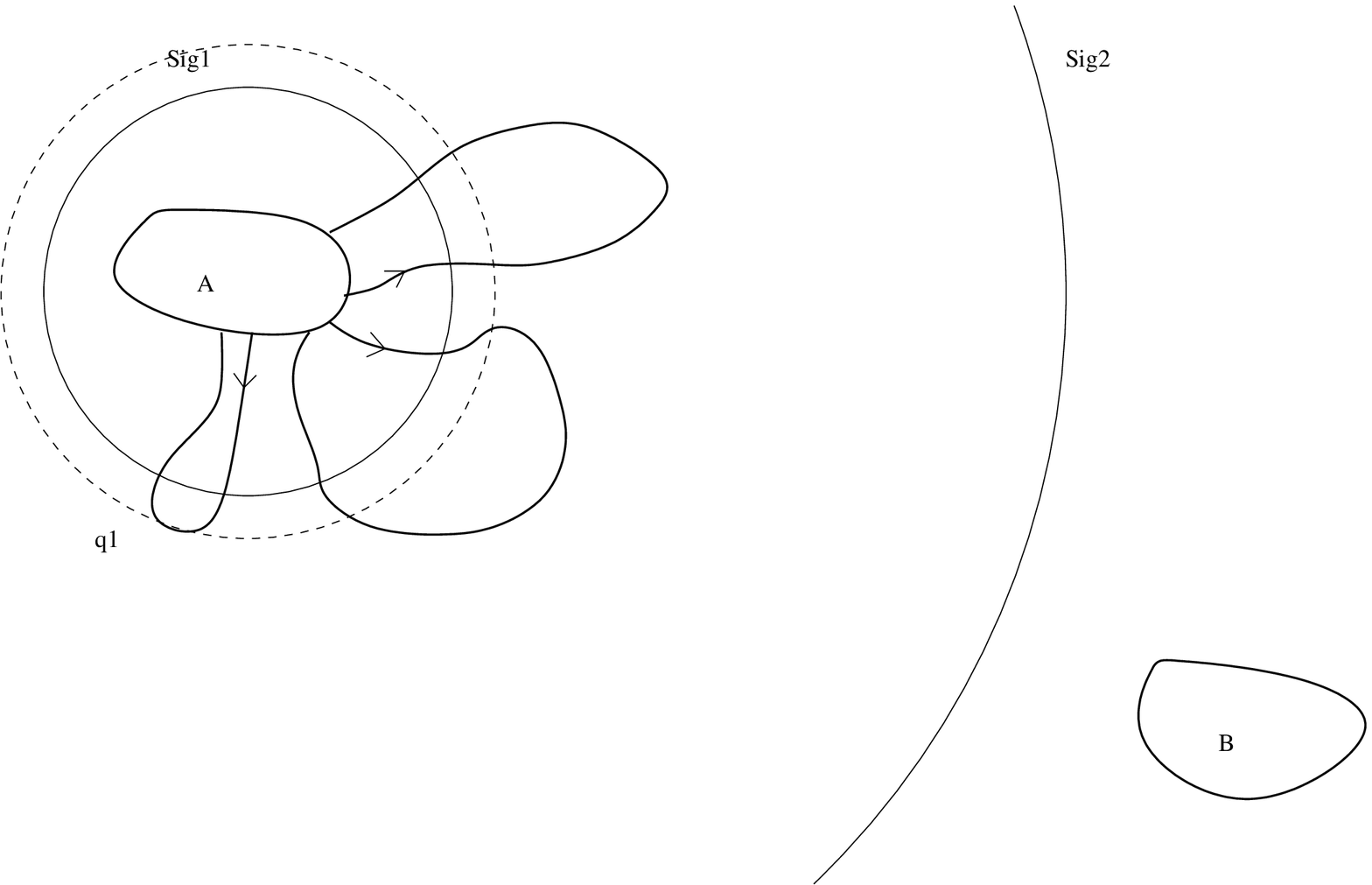}~~~~~~\includegraphics[width=6.5cm]{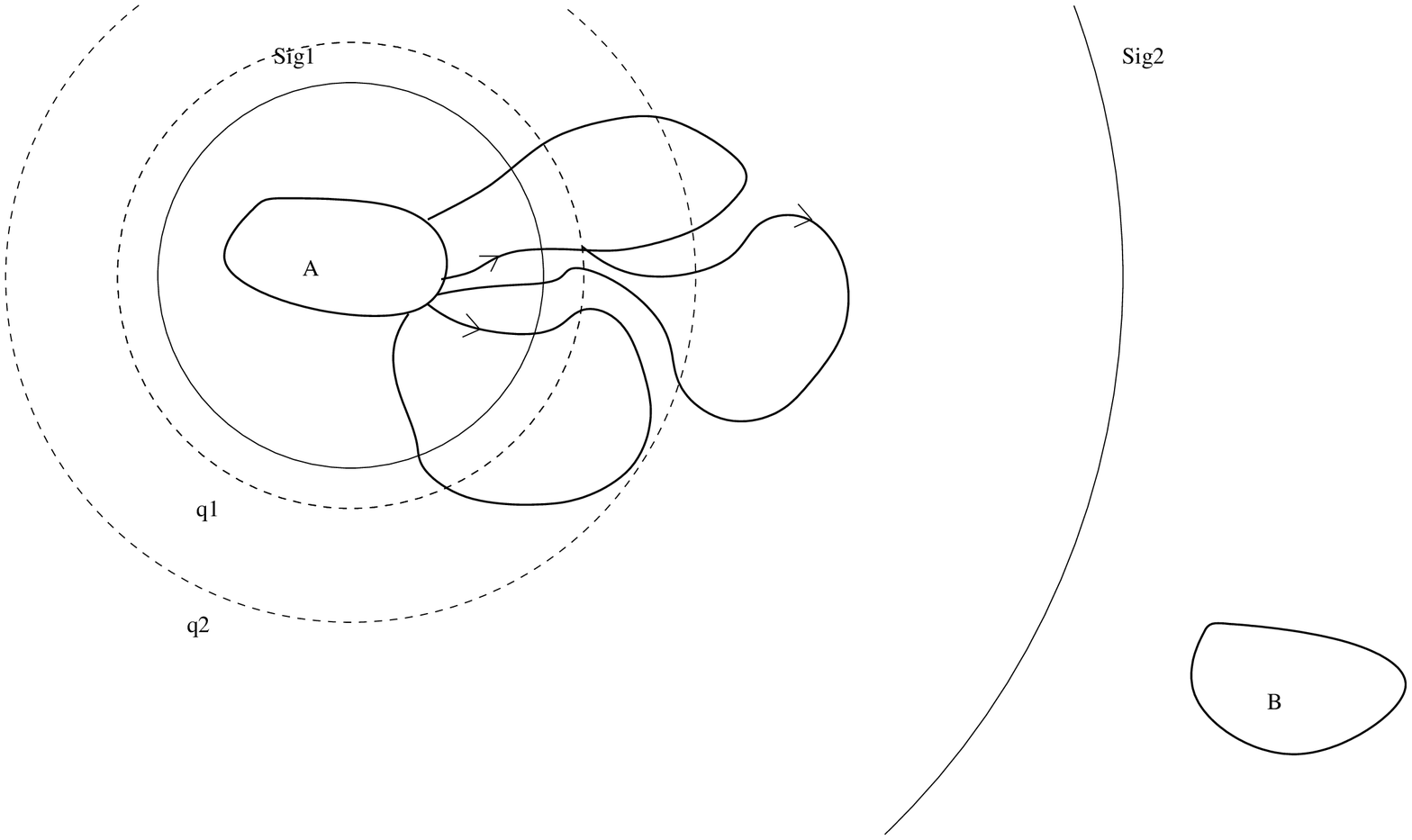}
  \caption{Schematic representation of the algorithm, with $N=3$ paths: on the left, the path which goes the less far in the reaction coordinate direction is killed, and, on the right, a new path is generated, starting from a previous one at the $1/N$ empirical quantile value.}
  \label{fig:algo}
\end{figure}

At the end of the initialization procedure, one has an ensemble of $N$ equilibrium trajectories $(X_t^n)_{0 \le t \le \tau^n}$, which leave $A$, end either in $A$ (the most likely) or in $B$, conditionally to the fact that $\sup_{t\ge 0} \xi(X^n_t) \ge z_{\rm min}$. Let us denote these trajectories $(X_t^{1,n})$ (for $n \in \{1,\ldots,N\}$) and the associated stopping times $\tau^{1,n}=\tau^n$. Now the algorithm goes as follows (see Figure~\ref{fig:algo} for a schematic representation): Iterate on $k \ge 1$,
\begin{enumerate}
\addtocounter{enumi}{3}
\item Compute the largest reaction coordinate value attained for each path:
$$z^{k,n}=\sup_{0 \le t \le \tau^{k,n}} \xi(X_t^{k,n}).$$
\item Order the values $(z^{k,n})_{1 \le n \le N}$:
$$z^{k,\varepsilon^k(1)} \le z^{k,\varepsilon^k(2)} \le \ldots \le z^{k,\varepsilon^k(N)},$$
where $\varepsilon^k$ is a permutation over $\{1,\ldots,N\}$. To simplify the notation, let us denote
$$n^k= \varepsilon^k(1)=\argmin_{n \in \{1, \ldots, N\}} z^{k,n},$$
the index which realizes the smallest value $z^{k,\varepsilon^k(1)}$, and let us denote $q^k$ the (empirical) quantile of order $1/N$, namely this smallest value:
 $$q^k=z^{k,\varepsilon^k(1)}=z^{k,n^k}=\min_{n \in \{1, \ldots, N\}} z^{k,n}.$$
\item Kill the trajectory $(X_t^{k,n^k})_{0 \le t \le \tau^{k,n^k}}$, and consider trajectories for iteration $k+1$ as follows:
\begin{itemize}
\item For all $n \neq n^k$, the $n$-th trajectory is unchanged: $\tau^{k+1,n}=\tau^{k,n}$ and $(X_t^{k+1,n})_{0 \le t \le \tau^{k+1,n}}=(X_t^{k,n})_{0 \le t \le \tau^{k,n}}$;
\item Generate a new $n^k$-th trajectory in three steps: (i) Choose at random (with uniform law) $i_k \in \{1, \ldots, n^k-1,n^k+1, \ldots, N\}$; (ii) Set $X_t^{k+1,n^k}=X_t^{k,i_k}$ for all $t \in (0,\sigma_k)$ where
$$\sigma_k= \inf \{t \ge 0, \xi(X_t^{k,i_k}) \ge q^k \};$$ (iii) Generate the end of the trajectory $(X_t^{k+1,n^k})_{t \ge \sigma_k}$ according to~\eqref{eq:sde} (with $W_t=W^{n_k}_t$) until the stopping time
$$\tau^{k+1,n_k}=\inf \{t \ge \sigma_k, X_t^{k+1,n^k} \in A \cup B \}.$$
\end{itemize}
\item Go back to 4 (with $k$ being $k+1$), until $q^k\ge z_{\rm max}$.
More precisely, the number of iterations is defined as
$$k_{\rm max}= \sup \{ k \ge 1, \, q^k \le z_{\rm max} \}.$$
\end{enumerate}
At iteration $k$ of the algorithm, one obtains an ensemble of $N$ equilibrium trajectories $(X_t^{k,n})_{0 \le t \le \tau^{k,n}}$, which leave $A$, end either in $A$ or in $B$, conditionally to the fact that $\sup_{ 0\leq t\leq \tau^{k,n}} \xi(X^n_t) \ge q^k$.

At the end of the algorithm,  all the trajectories cross the submanifold $\Sigma_{q^{k_{\rm max}}}$. Since $k_{\rm max}$ is the last iteration index for which the quantile $q^k$ is smaller than $z_{\rm max}$ and since $\Sigma_{z_{\rm max}}$ is assumed to be ``close to''~$B$, most of them end in~$B$. The final step to retain only reactive trajectories is:
\begin{enumerate}
\addtocounter{enumi}{7}
\item We retain only the trajectories which indeed end in $B$ to perform statistics on reactive trajectories. We denote $r$ the proportion of such trajectories among the ones obtained at the final iteration $k_{\rm max}$.
\end{enumerate}
This last step does not introduce any additional bias since (recall that $q^{k_{\rm max}} \le z_{\rm max}$)
$$B \subset \{ x \in \R^d , \, \xi(x) \ge q^{k_{\rm max}} \}.$$

Note that it is very simple to adapt this algorithm to any stochastic Markovian dynamics, the new paths being generated at each iteration using independently drawn random numbers.

\subsection{Discussion of the algorithm}
As mentioned above, at the end of the $k$-th step of the algorithm (steps 4 to 7), one has an ensemble of $N$ equilibrium trajectories, which start from $A$, end either in $A$ or $B$, and are distributed conditionally to the event $\{\sup_{0 \le t \le \tau} \xi(X_t) \ge q^k\}$. It can be shown (see~\cite{cerou-guyader-07}) that the bias scales like $O(1/N)$ and that the standard deviation scales like $O(1/\sqrt{N})$. 

At the end of the day, note that
\begin{equation}
  \label{eq:alphaN}
\hat{\alpha}_N=r\left(1-\frac{1}{N}\right)^{k_{\rm max}}  
\end{equation}
gives an estimate of the probability $p$ of ``observing a reactive trajectory'', since at each iteration of the algorithm, only one in the $N$ trajectories is killed. More precisely, this is an estimate of the probability, starting from $\Sigma_{z_{\rm min}}$ with the equilibrium distribution generated by the initialization procedure (steps~1 to~3 above), to observe a trajectory which touches $B$ before $A$. This probability actually depends on the choice of $\Sigma_{z_{\rm min}}$, while the law of the reactive trajectories generated by the algorithm does not.

There exist generalizations of the above algorithm (see~\cite{cerou-guyader-07}), where, instead of killing only $1$ trajectory and working with the quantile of order $\frac{1}{N}$, $i$ trajectories are killed and the quantile of order $\frac{i}{N}$ is considered (where $i \in \{1, \ldots,N-1\}$). We observe numerically (and this can be checked theoretically at least in some simple situations, see~\cite{guyader-hengartner-matzner-10}) that $i=1$ yields the best results. However, in those preliminary simulations, we did not use the fact that the algorithm with a quantile of order $\frac{i}{N}$ (and $i>1$) enables to simulate the new $i$ replicas in parallel, which may lead to an interesting computational gain in cases when simulating a new trajectory is  computationally demanding.

A difficult mostly open question is to compute the asymptotic variance of an estimator associated to a given observable over reactive trajectories (say the time length of the trajectory) and then to try to optimize this estimator with respect to the chosen reaction coordinate. It can be shown that in terms of the asymptotic variance of $\hat{\alpha}_N$, the optimal reaction coordinate (commonly called {\em importance function} in the context of statistics, see~\cite{dean-dupuis-09}) is the so-called {\em committor function} $q$ (see~\cite{hummer-04,e-vanden-eijnden-04}) which satisfies:
\begin{equation}\label{eq:committor}
\left\{
\begin{array}{l}
- \nabla V \cdot \nabla q + \beta^{-1} \Delta q = 0 \text{ in $\R^d \setminus (\overline{A} \cup \overline{B})$},\\[3pt]
q=0 \text{ on $\partial A$ and } q=1 \text{ on $\partial B$}.
\end{array}
\right.
\end{equation}
The function $q$ can be interpreted in terms of the process $X_t^x$ solution to~\eqref{eq:sde} with initial condition $X_0^x=x$, as:
\begin{equation}\label{eq:committor_interpretation}
q(x)=\P \left(X_t^x \text{ reaches $B$ before $A$} \right).
\end{equation}
We will check numerically below that the variance of the results seems to be smaller if $\xi$ is chosen close to $q$. But one interesting feature of the method is that it does not need to be the case to give reliable results in terms of reactive trajectories.

\subsection{Complexity of the algorithm}

The number of iterations $k_{\rm max}$ of course depends on the problem at
hand.  If the probability of reaching  $\{x \in \R^d, \xi(x) \ge z_{\rm max}
\}$ starting at equilibrium from $\Sigma_{z_{\rm min}}$
without touching $A$ is denoted $\overline{p}$, then it has been proved in~\cite{guyader-hengartner-matzner-10} that the random variable $k_{\rm max}$ has a Poisson distribution, with mean $\E[k_{\rm max}]=-N \log ( \overline{p} )$. In the numerical simulations below, the probability $\overline{p}$ ranges between $10^{-18}$ and $10^{-4}$ depending on the situation considered.

The initialization procedure (steps 1 to 3) is straightforward to parallelize. Then, note that at each iteration, only one new trajectory needs to be computed, until it reaches one of the two metastable states $A$ or $B$. Since $A$ and $B$ are typically metastable sets that equilibrium trajectories visit very often, this should not be too costly. Moreover, inserting the new maximum value $z^{k+1,n^k}$ within the set of former values is a very rapid procedure, since they were already ordered. In practice, we use a dichotomy recursive algorithm, by comparing $z^{k+1,n^k}$ to the median value (quantile or order $1/2$) of $\{z^{k,1}, \ldots,z^{k,n^k-1},z^{k,n^k+1}, \ldots, z^{k,N}\}$, and then to the quantile of order $1/4$ or $3/4$ depending on its position with respect to the median value, etc. Then the insertion of the new particle at the right 
place in the ordered sample has also a cost in $O(\log N)$ via a min-heap algorithm (see for 
example \cite{knuth3}).  This yields an algorithm of complexity $O(\log N)$ at each iteration.

Concerning the cost of the algorithm in terms of memory, two remarks are in order. First, for trajectories which end in $A$ rather than $B$, one can discard the part of the trajectory after it reaches its maximum value along $\xi$, since this piece is then never used in the algorithm. Second, at iteration $k$, one may only retain a coarse description of each trajectory up to the first time it reaches the quantile value $q^k$ along $\xi$. Indeed, in the forthcoming iterations, this part of the trajectory is not used anymore. By coarse description, we mean that one only needs to retain the features of that part of the trajectory that are needed to perform the required statistics over reactive trajectories. For example, if only the time length of the reactive trajectory is needed (see for example the results in Section~\ref{sec:num}), the whole trajectory until this time may be discarded. In any case, one could think of storing only a subsampling of this part of the trajectory.

\section{Computing reactive trajectories: numerical illustrations}
\label{sec:num}

\subsection{A one-dimensional case}
\label{sec:1d}

In this section, we consider a one-dimensional situation and overdamped Langevin trajectories~\eqref{eq:sde}, with $V$ being the double-well potential:
$$V(x)=x^4 -2 x^2.$$
This potential has two global minima at $\pm 1$ and one local maximum at $0$.
In this simple one dimensional setting, we set as metastable states $A=\{-1\}$ and $B=\{+1\}$, and the reaction coordinate is taken to be simply
$$\xi(x)=x.$$
For the numerical experiments, we take $z_{\rm {max}}=-z_{\rm {min}}=0.9$. The aim of this section is mainly to validate the algorithm by comparing the results to those obtained by direct numerical simulation (DNS), namely a simple Monte Carlo algorithm without any selection procedure, and then to have an idea of the computational gain. DNS can only be used for small values of $\beta$ (typically $\beta \le 10$ in this setting).

\paragraph{Distribution of the time lengths of reactive paths.}

\begin{figure}[htbp]
  \centering
   \epsfig{file=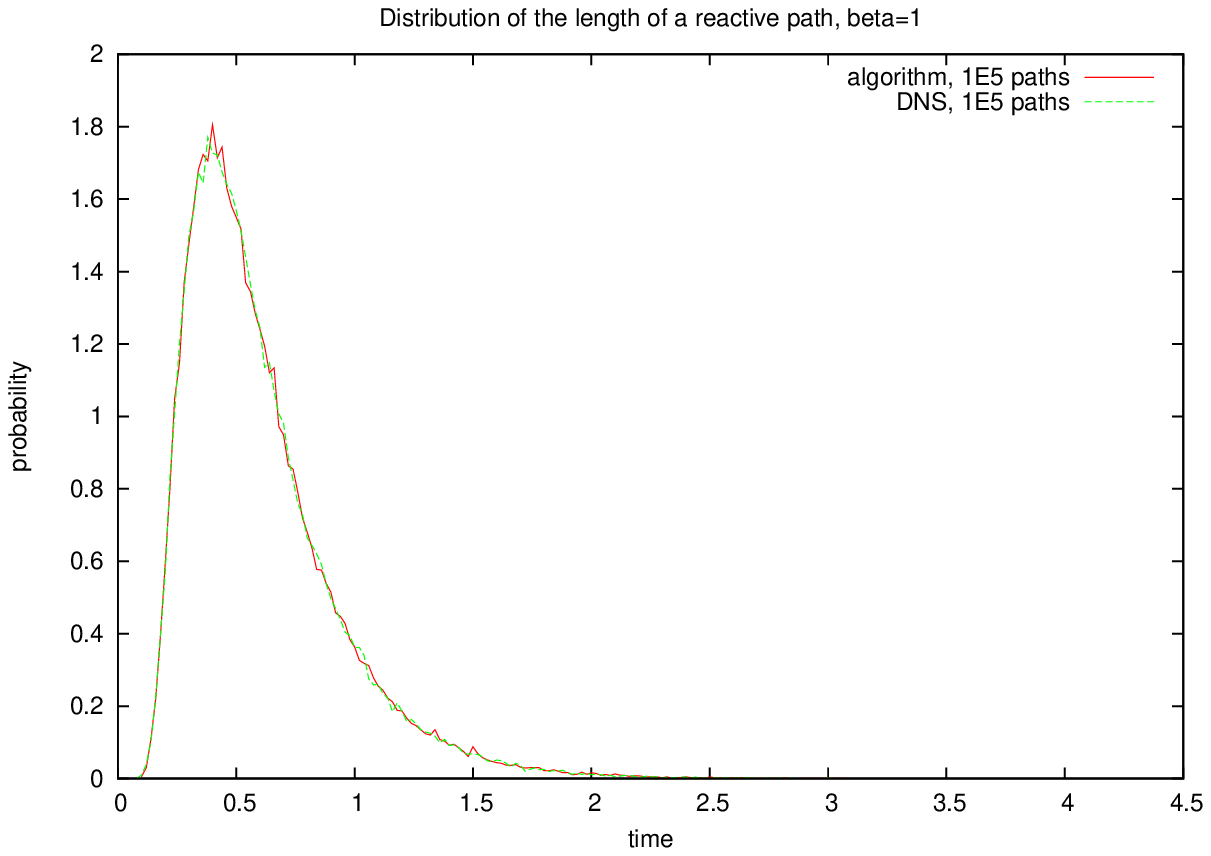,width=6cm}
  \epsfig{file=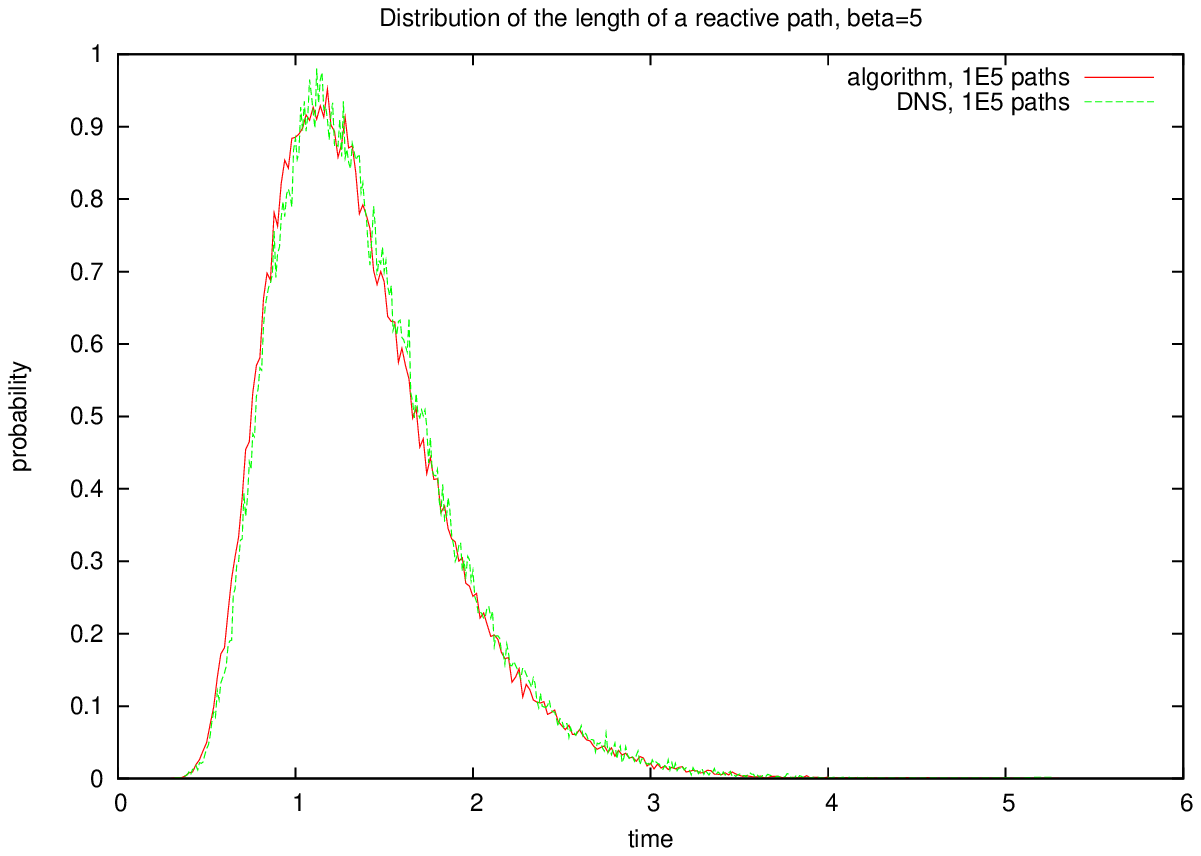,width=6cm}
  \epsfig{file=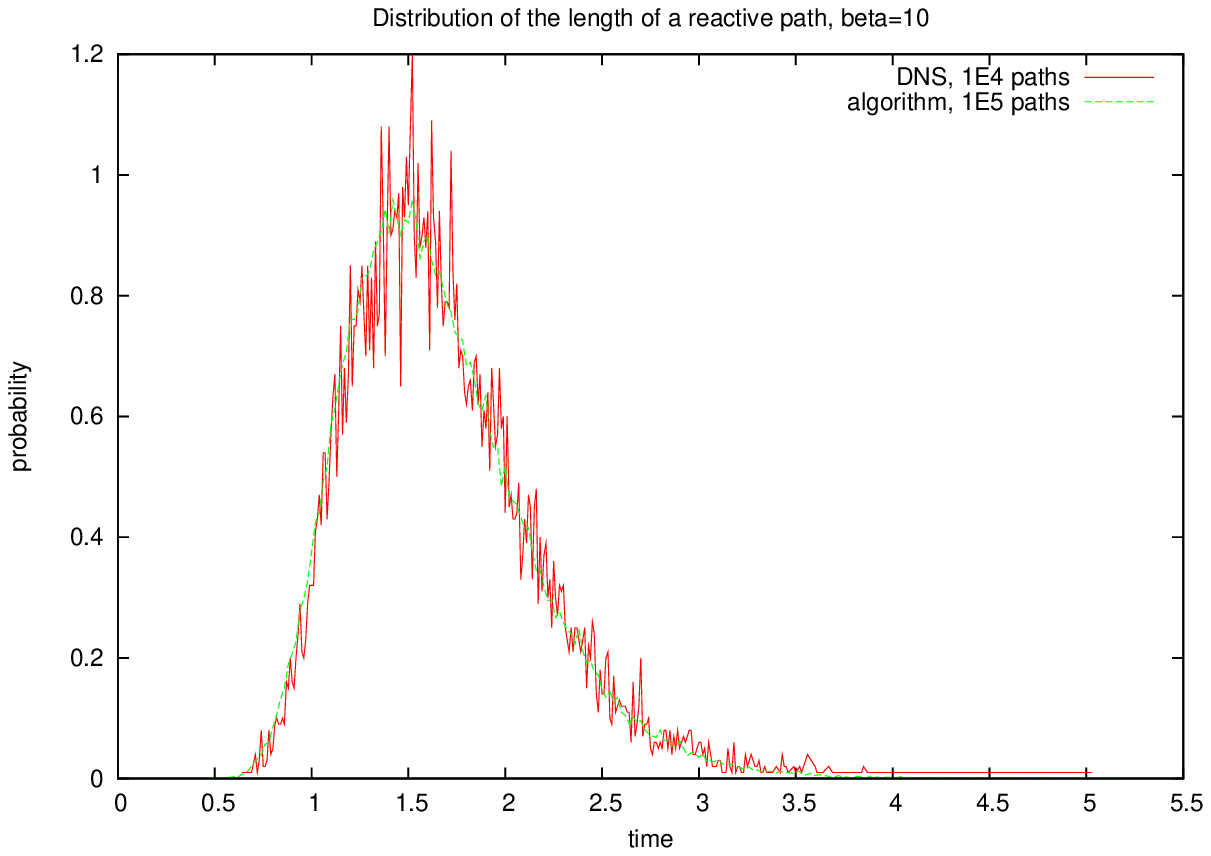,width=6cm}
  \epsfig{file=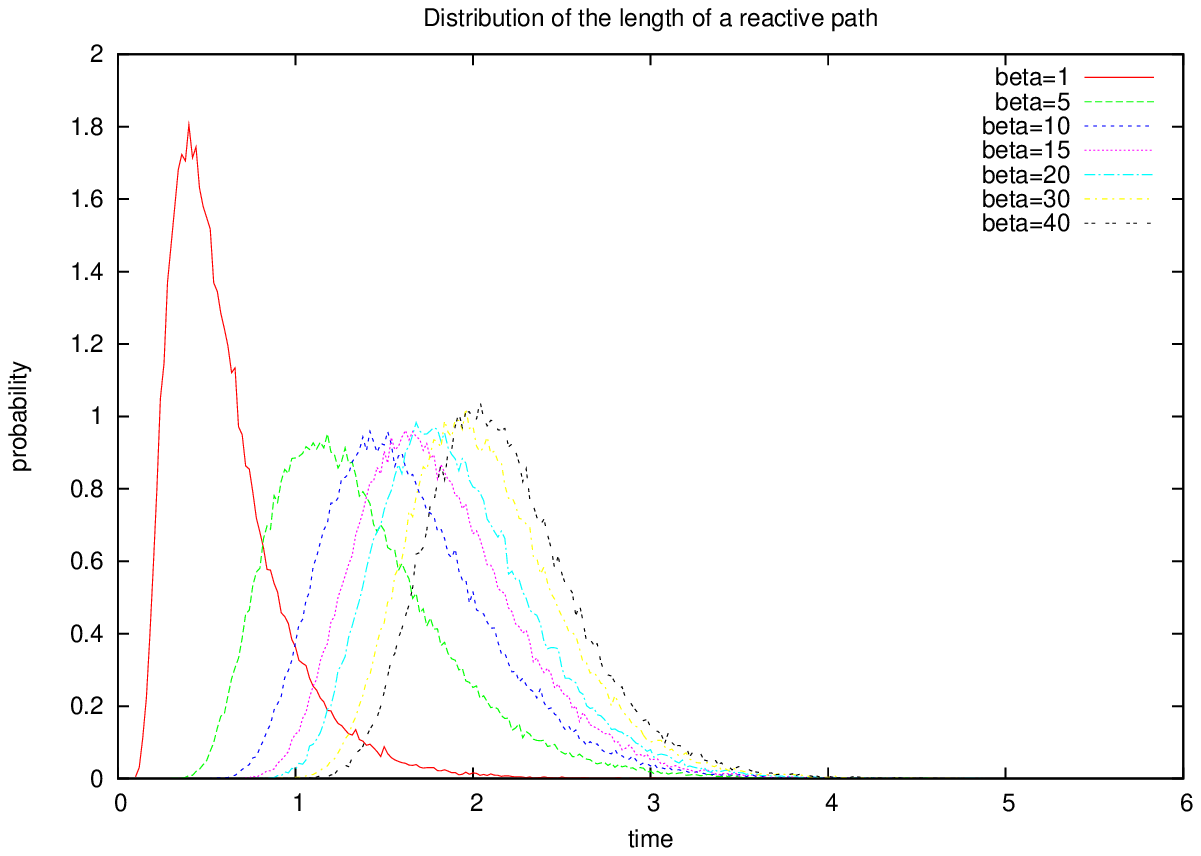,width=6cm}
  \caption{Distribution of the time lengths of reactive paths. In the first three figures we
   compare results computed with DNS and our algorithm. In the last figure, distributions of the lengths are given for various values of $\beta$:  $\beta=1,5,10,15,20,30,40$.}
  \label{fig:time1d}
\end{figure}

To validate the algorithm, we compute an histogram of the distribution of the time length (duration) of a reactive path. On Figure~\ref{fig:time1d}, we compare the results obtained with DNS and our algorithm: the agreement is excellent. The interest of our algorithm is that it is possible to compute this distribution for very small temperatures (large values of $\beta$), for which a DNS cannot be used.

\begin{figure}[htbp]
  \centering
  \epsfig{file=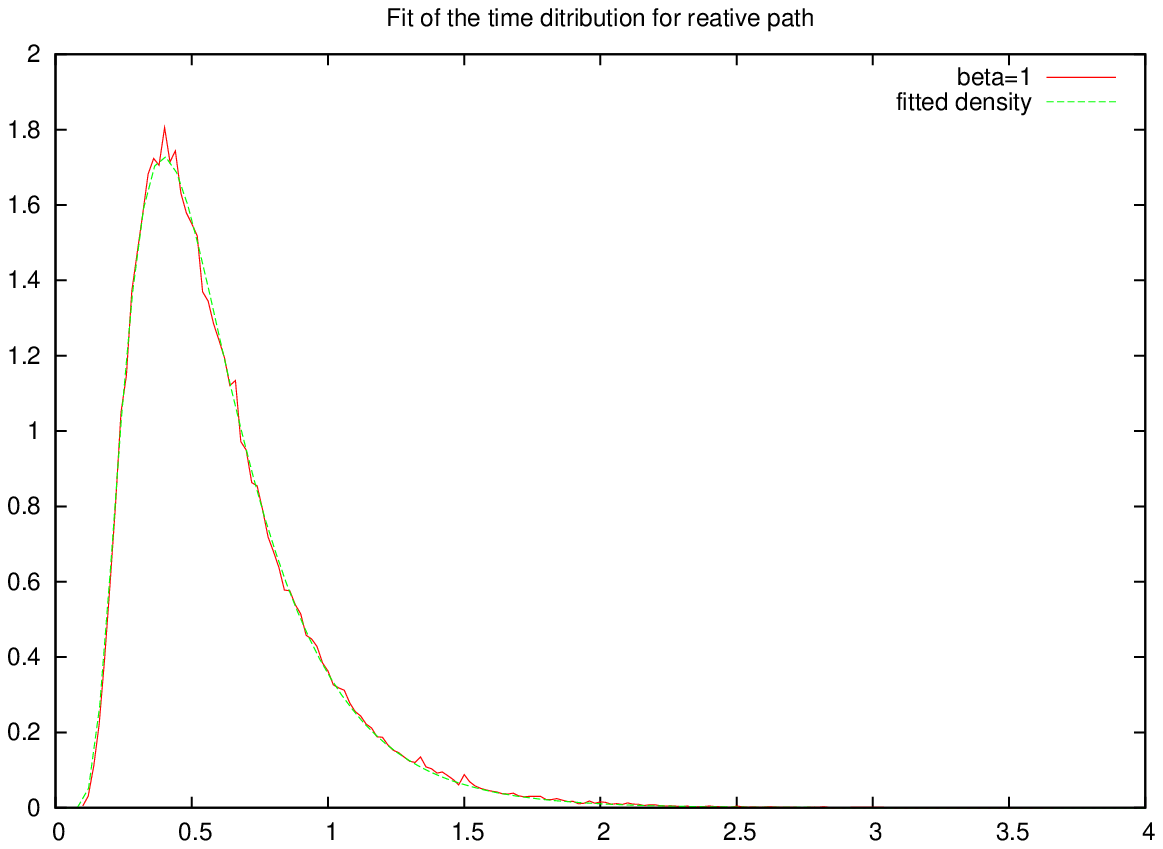,width=6cm}
  \epsfig{file=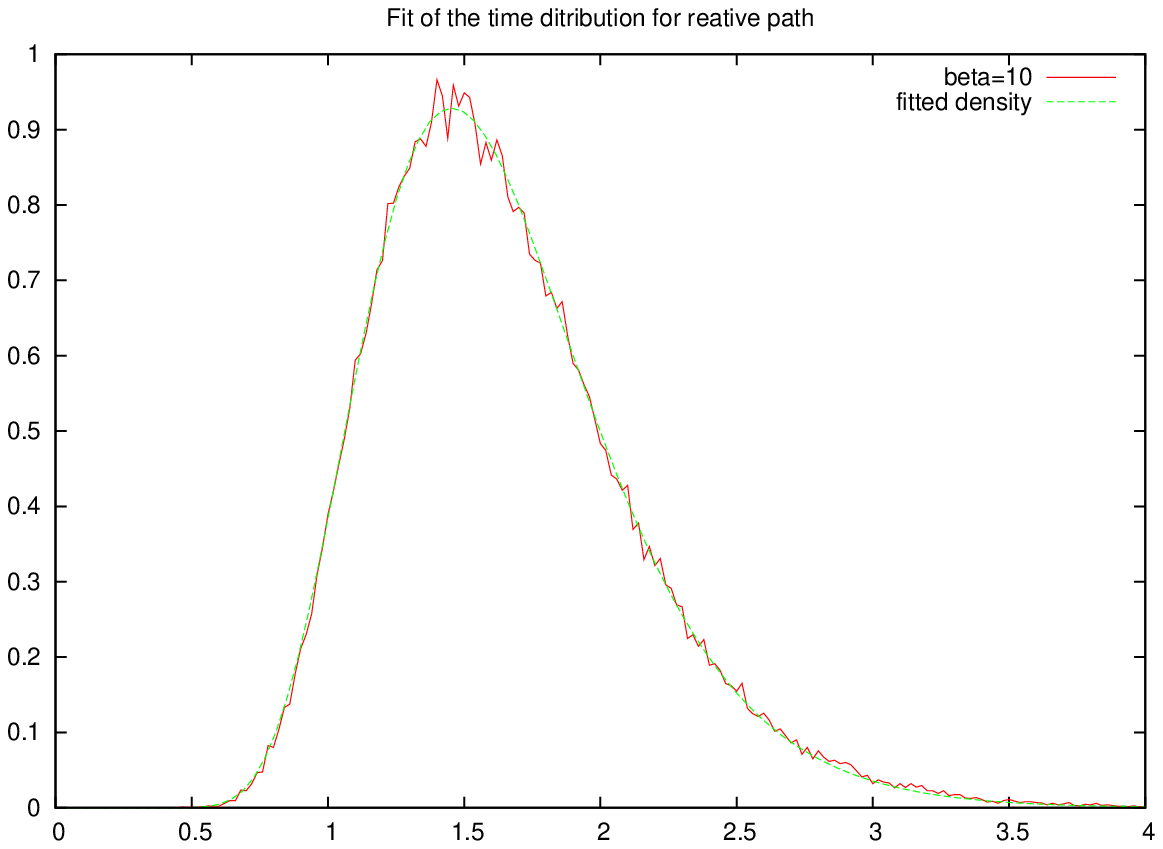,width=6cm}
  \epsfig{file=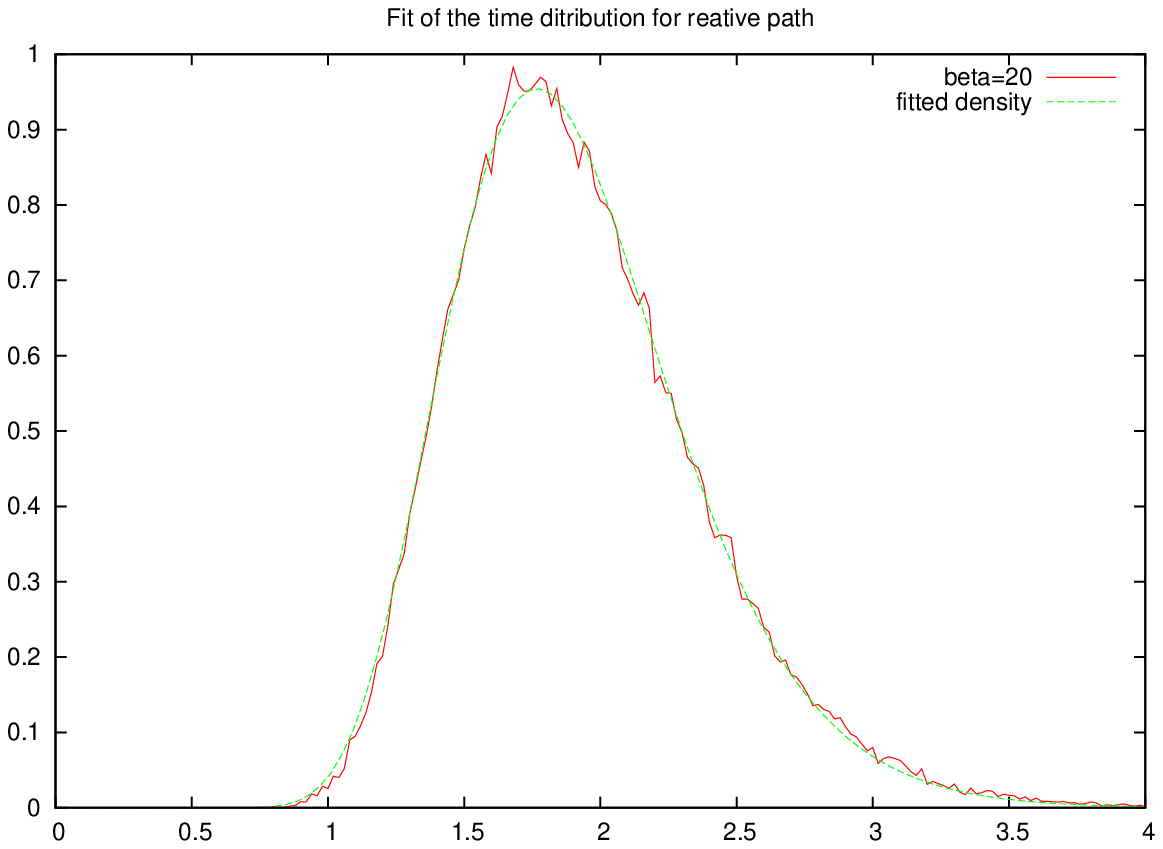,width=6cm}
  \epsfig{file=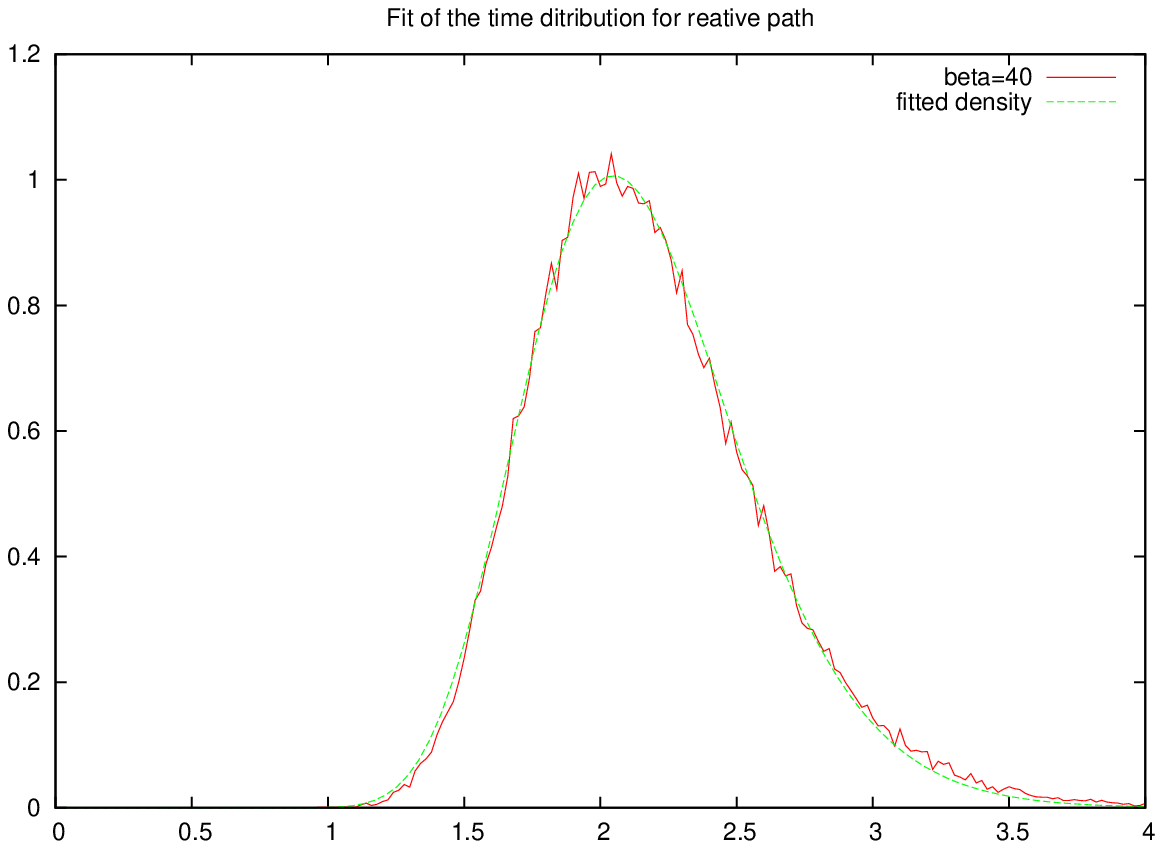,width=6cm}
 \caption{Distribution of the time lengths of reactive paths and the inverse
   Gaussian law fitted to this distribution, for $\beta=1,10,20,40$.}
  \label{fig:error_fit1d}
\end{figure}

We observe (see Figure~\ref{fig:error_fit1d}) that this distribution seems to
be very close to an inverse Gaussian distribution with density (see also~\cite{luccioli2010unf}):
$$f_{\mu,\lambda}(x) = \left(\frac{\lambda}{2\pi x^3} \right)^{1/2}\exp{\left(\frac{- \lambda
  (x-\mu)^2}{2\mu^2 x}\right)} \un_{x > 0},$$ 
where $\mu$ and $\lambda$ are two positive parameters. Inverse Gaussian distributions naturally appear when considering first passage time at a fixed level $\alpha > 0$ for a one-dimensional stochastic process $Y_t= \nu t + \sigma W_t$:
$$    T_{\alpha} = \inf\{ t>0 \mid Y_t=\alpha \} \sim IG \left(\tfrac\alpha\nu, \tfrac {\alpha^2} {\sigma^2}\right).$$ 
Optimal values for the parameters $(\mu,\lambda)$ to fit the time lengths distributions are provided in
Table~\ref{tab:1d_fit}. We again observe that the agreement between DNS and our algorithm is excellent.

\begin{table}
  \centering
  \begin{tabular}{|c|c|c|c|c|c|} \hline
    $\beta$&DNS $(\mu ,\lambda)$& Algo $(\mu , \lambda)$\\[2pt] \hline
    $1$   & $0.59 , 2.14 $ & $0.60 , 2.14$ \\ 
    $5$   & $1.35, 10.34 $ & $1.36 , 10.38$\\ 
    $10$ & & $1.64 , 20.02$ \\ 
    $15$ &  & $1.80 , 28.40$\\ 
    $20$ &  & $1.91 , 35.63$\\ 
   $40$ &  & $2.16, 59.32$\\ \hline
  \end{tabular}
 \caption{ Fitted parameter on Inverse Gaussian law of the probability distribution of the time stopping of reactive path for various
  values of $\beta$.}  \label{tab:1d_fit}
\end{table}

\paragraph{Computational time.}

Let us now compare the computational time required to simulate an ensemble of
reactive paths. All the computations have been performed using an Intel Xeon
CPU ($2.50GHz$), with $16$ Go RAM. We use a Mersenne Twister algorithm~\cite{matsumoto-nishimura-98}
for the random number generator.

In Table~\ref{tab:1d_res}, we give CPU times for various values of $\beta$, using DNS (when possible) or our algorithm. The DNS time simulation rapidly explodes when $\beta$ increases. For $\beta=15$ and $N=10^5$, the ratio between the CPU time of a DNS and the CPU time of our algorithm is of the order of $1\,000$.

\paragraph{Variance of the estimators $\hat{\alpha}_N$.}

To complete the discussion on computational time, we also compare in Table~\ref{tab:1d_res} the relative variances of the estimators  $\hat{\alpha}_N$ of the probability $p$, for DNS and for our algorithm. The relative variance is defined as the variance divided by the mean square. With the notations of this table, the relative variance of the DNS estimator for $\hat{\alpha}_N$ is estimated by $(1-\hat{\alpha}_N)/N$. With our algorithm, it has been proved \cite{guyader-hengartner-matzner-10} that this relative variance can be estimated by $-\log(\hat{\alpha}_N)/N$. This explains why in the five last rows of Table~\ref{tab:1d_res} (where $N=10^{5}$), the relative variance of our estimator increases very slowly (in fact, logarithmically) when the probability of interest decreases to zero. To take into account both computational time and variance, it has been proposed in~\cite{hammersley} to define the efficiency of a Monte Carlo process as ``inversely proportional to the product of the sampling variance and the amount of labour expended in obtaining this estimate.'' Using this definition of efficiency, for $\beta=15$ and $N=10^5$, our algorithm is about 800 times more efficient than DNS.

\begin{table}
  \centering
  \begin{tabular}{|c|c|c|c|c|c|c|c|} \hline
    $\beta$& $N$& $\hat{\alpha}_N$& DNS CPU
    & Algo CPU & DNS RV & Algo RV \\ \hline
    $1$   & $10^4$ & $1.03 \, 10^{-1}$ &$2s$&$2s$&$9 \, 10^{-5}$& $2 \, 10^{-4}$\\ 
    $1$   & $10^5$ & $1.01 \, 10^{-1}$ &$21s$&$ 1 \text{ min } 19 \text{ s }$&$9 \, 10^{-6}$&$ 2 \, 10^{-5}$  \\ 
    $10$ & $10^4$ & $2.04 \, 10^{-5}$ & $140 \text{ min } 05 \text{ s }$&$ 5 \text{ s }$ &$10^{-4}$& $10^{-3}$\\ 
    $10$ & $10^5$ & $1.98 \, 10^{-5}$ & $1400 \text{ min }^{\star}$&$5 \text{ min } 22 \text{ s }$&$10^{-5}$& $10^{-4}$ \\ 
    $15$ & $10^5$ & $1.78 \, 10^{-7}$ & $92000 \text{ min }^{\star}$&$7 \text{ min } 52 \text{ s }$&$10^{-5}$& $1.5 \, 10^{-4}$ \\ 
    $20$ & $10^5$ & $1.33 \, 10^{-9}$ & &$8 \text{ min }36 \text{ s }$&& $2 \, 10^{-4}$ \\ 
    $40$ & $10^5$ & $5.82 \, 10^{-18}$ & &$10 \text{ min } 09 \text{ s }$&& $4 \, 10^{-4}$ \\ \hline
\end{tabular}
  \caption{Probability $\hat{\alpha}_N$ (see~\eqref{eq:alphaN}), computational time and relative variance (RV) for the estimators of $p$, for different
  values of~$\beta$ and number of paths $N$. DNS CPU
  time with $\star$ is an extrapolated time deduced from a small number of
  iterations.}
 \label{tab:1d_res}
\end{table}

\subsection{A two-dimensional case with two channels}
\label{sec:2d}

\begin{figure}[htbp]
  \centering
\epsfig{file=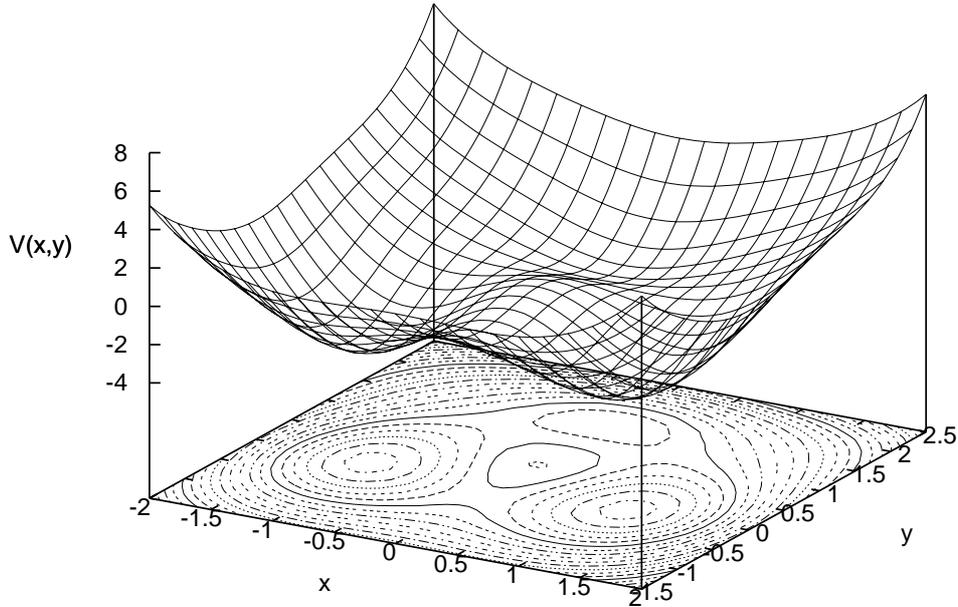,width=8cm,angle=270}
 \caption{The potential $V$.}
  \label{fig:V}
\end{figure}

In this section, we apply the algorithm to a two-dimensional situation, again with overdamped Langevin trajectories~\eqref{eq:sde}. The potential $V$ we consider is taken from~\cite{park-sener-lu-schulten-03,metzner-schuette-vanden-eijnden-06}:
 \begin{equation}
\begin{aligned}
   V(x,y)& = 3 {\rm e}^{-x^2 -(y-\frac{1}{3})^2} - 3 {\rm e}^{-x^2
     -(y-\frac{5}{3})^2} - 5 {\rm e}^{-(x-1)^2  -y^2}\\
& \quad- 5 {\rm e}^{-(x+1)^2 -y^2}+ 0.2 x^4 + 0.2 \left(y - \frac{1}{3}\right)^4.
 \end{aligned}
\end{equation}
 This potential (see Figure~\ref{fig:V}) has two deep minima approximately at $H_{\pm}=(\pm1,0)$, a
 shallow minimum approximately at $M = (0,1.5)$ and three saddle points
 approximately at $U_{\pm}= (\pm 0.6,1.1)$ and $L=(0,-0.4)$. In the notation of our algorithm above, $A$ (resp. $B$) denotes a neighborhood of $H_{-}$ (resp. $H_+$). The two metastable regions $A$ and $B$ can thus be connected by two channels: The upper channel around the points
 $(H_{-},U_{-},M,U_{+},H_{+})$ and the lower channel around the points $(H_{-},L,H_{+})$.

{F}rom large deviation theory~\cite{freidlin-wentzell-84}, it is known that in the small temperature limit ($\beta \to \infty$) the
reactive trajectories which will be favored will go through the upper channel, since the
energy barrier is lower there. On the other hand, at higher temperature, the lower channel is also very likely, since the trajectories going through the upper channel visit a broader region of space, and are thus less favored due to an entropic effect. In other words, the lower channel is more narrow. This entropic switching effect is well-known and has been studied in~\cite{park-sener-lu-schulten-03,metzner-schuette-vanden-eijnden-06}.

For the numerical experiments, following~\cite{metzner-schuette-vanden-eijnden-06}, we use the two following values for the
temperature: $\beta = 6.67$ (low temperature), which is such that most of the
trajectories go through the upper channel, and $\beta =1.67$ (high temperature),
which is such that most of the trajectories go through the lower channel.

The region $A$ (resp. $B$) is defined as the Euclidean ball ${\mathcal B}(H_-,0.05)$ (resp. ${\mathcal B}(H_+,0.05)$). To discretize the dynamics~\eqref{eq:sde}, we use an Euler scheme with a time-step size $\Delta t=10^{-2}$.
To test the influence of the choice of the reaction coordinate, we will consider two different reaction coordinates:
\begin{equation}\label{eq:2RC}
\xi_1(x,y)=x, \text{ and } \xi_2(x,y)=\|(x,y)-H_-\|.
\end{equation}

For the numerical experiments, we take $z_{\rm {min}}=0.05$ (resp. $z_{\rm {max}}=0.9$)
when the reaction coordinates is $\xi_1$ and $z_{\rm {min}}=0.05$ (resp.
$z_{\rm {max}}=1.5$) when the reaction coordinates is $\xi_2$. We will plot
the probability density~$\rho$ of positions, conditionally on being on a
reactive trajectory. The discretization of this density uses a regular grid of
size $100 \times 100$ with constant $x$ and $y$ intervals.

\begin{figure}
  \centering
  \epsfig{file=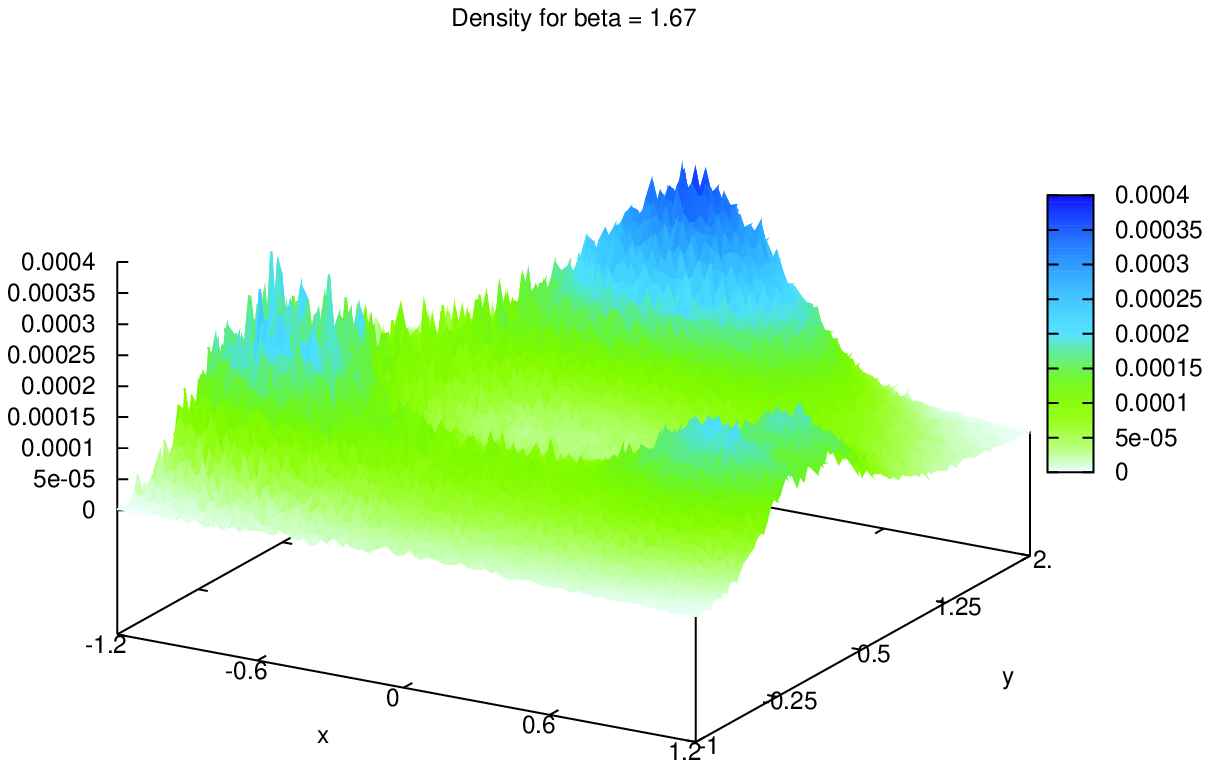,width=6cm}
 \epsfig{file=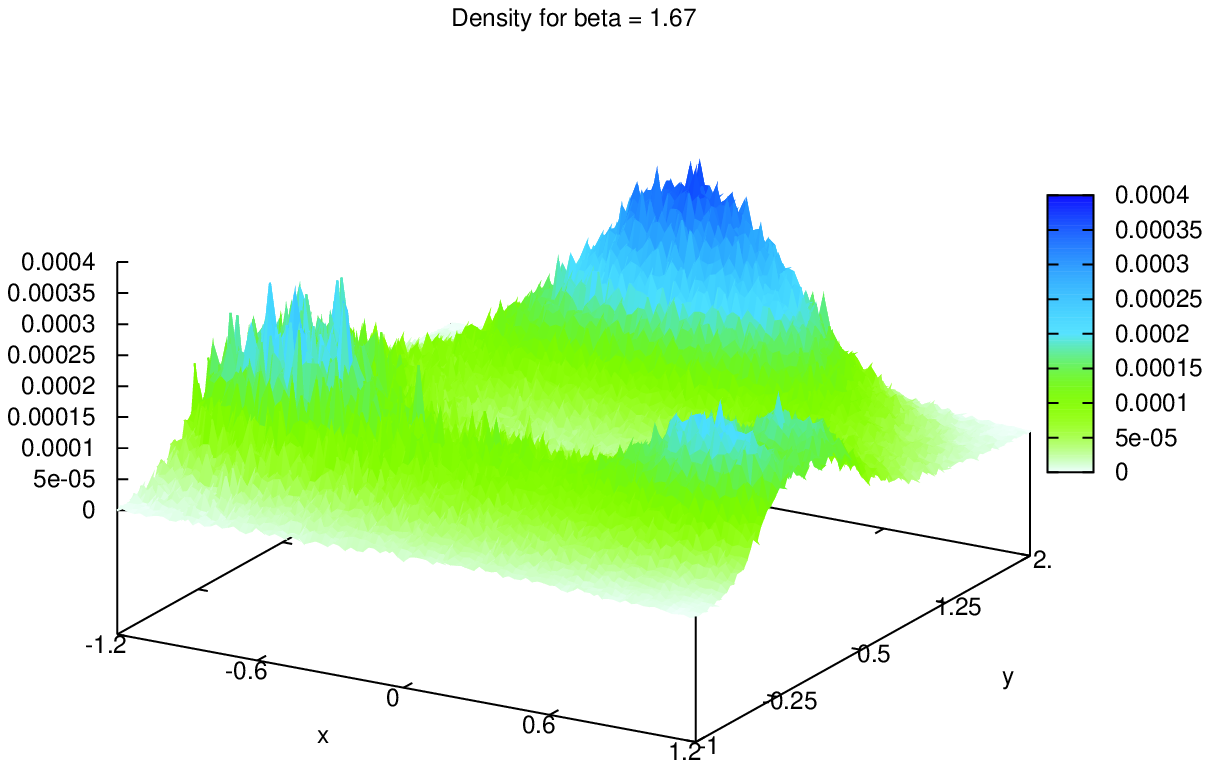,width=6cm}
 \epsfig{file=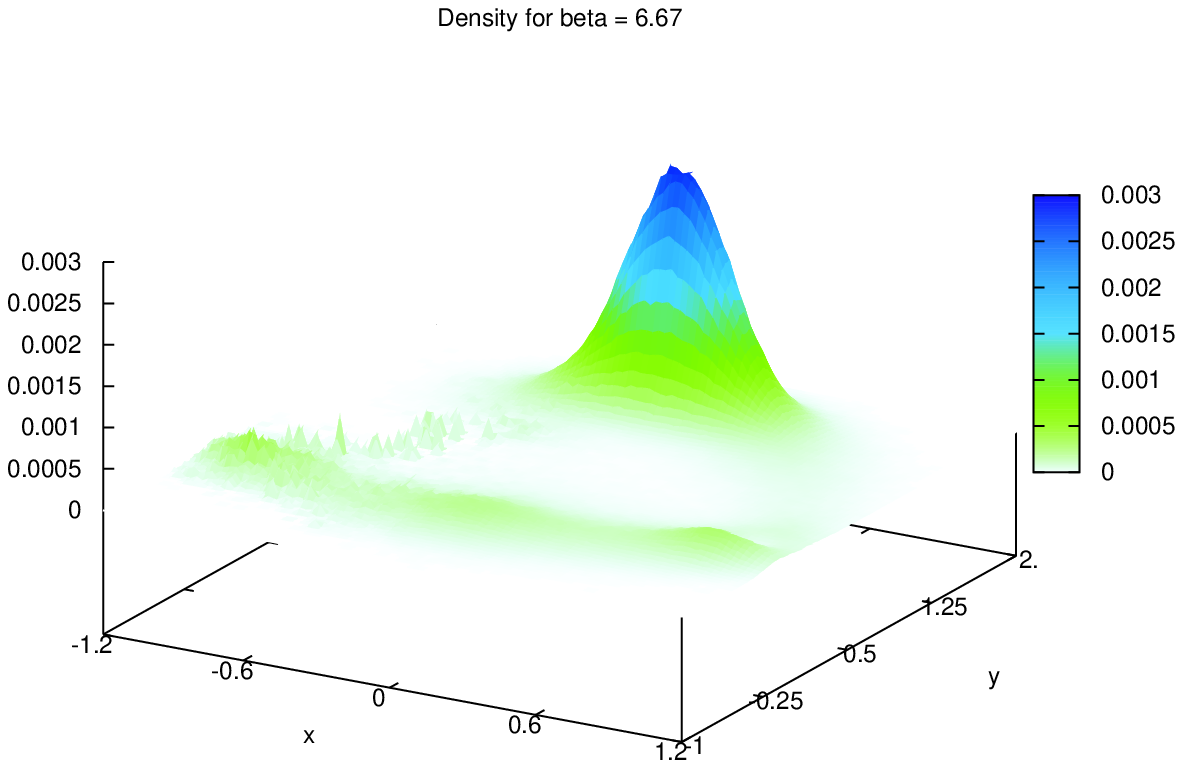,width=6cm}
 \epsfig{file=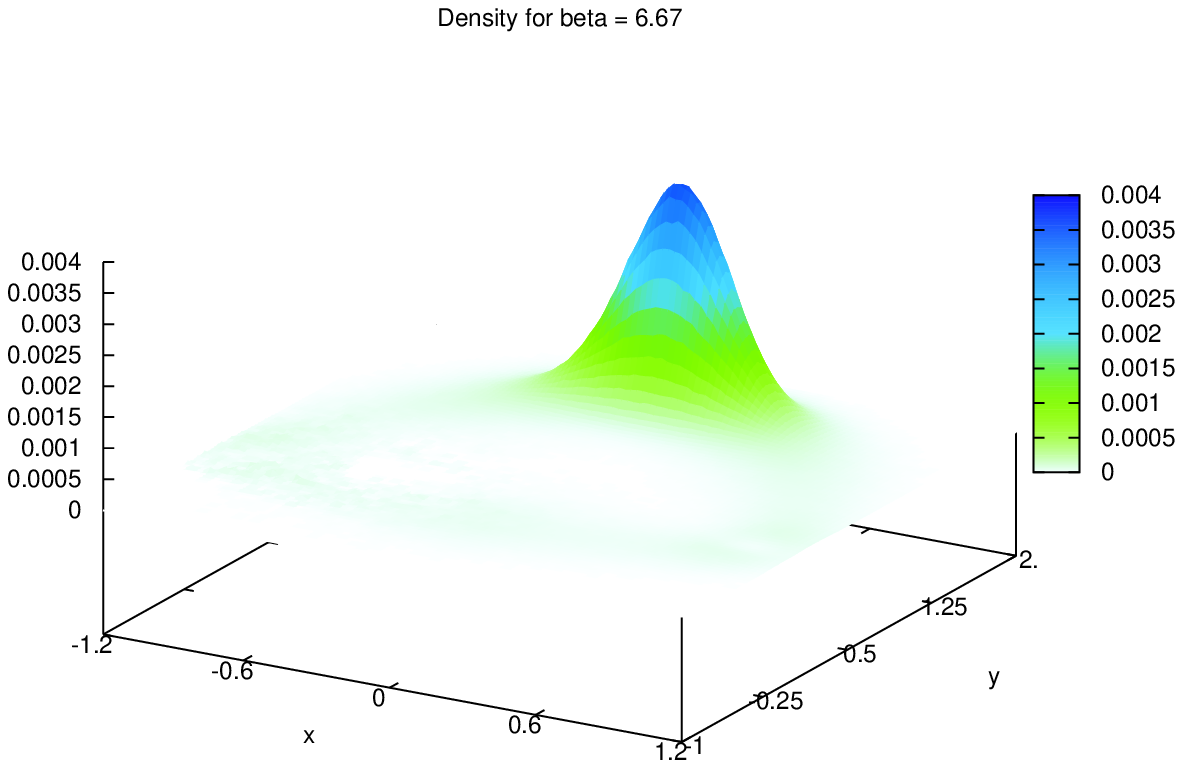,width=6cm}
 \epsfig{file=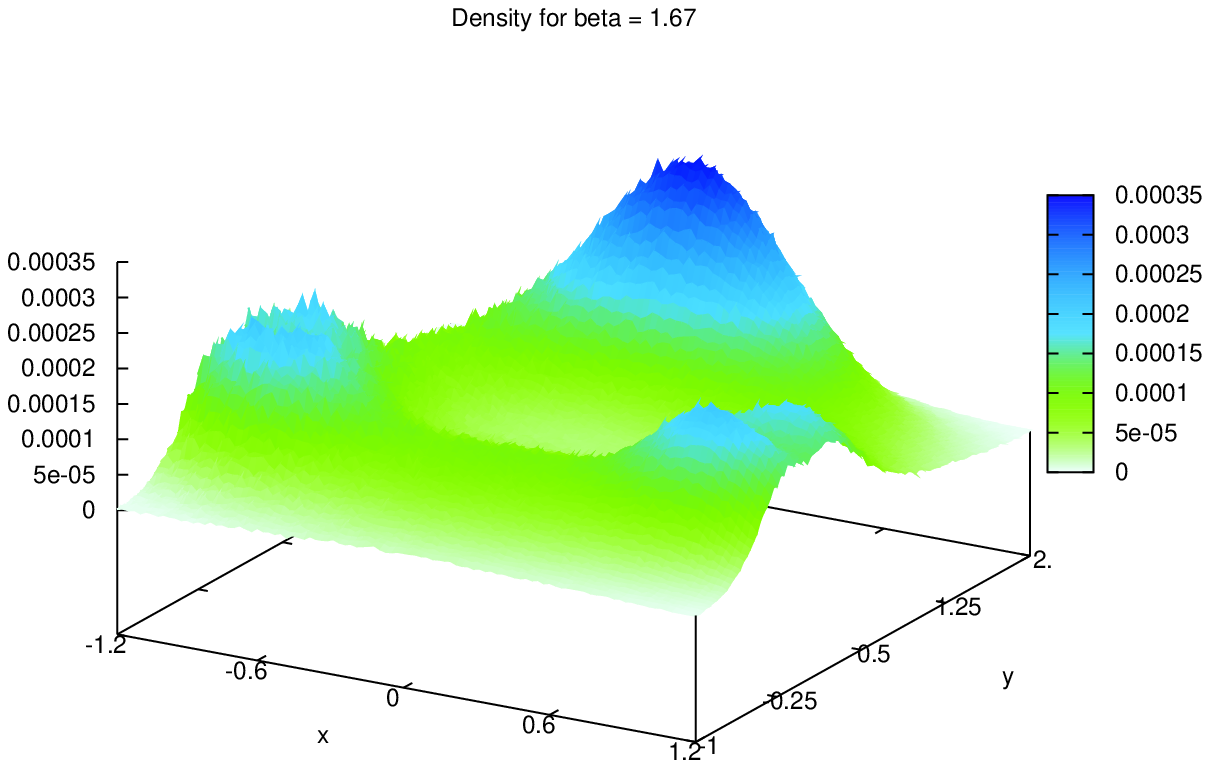,width=6cm}
 \epsfig{file=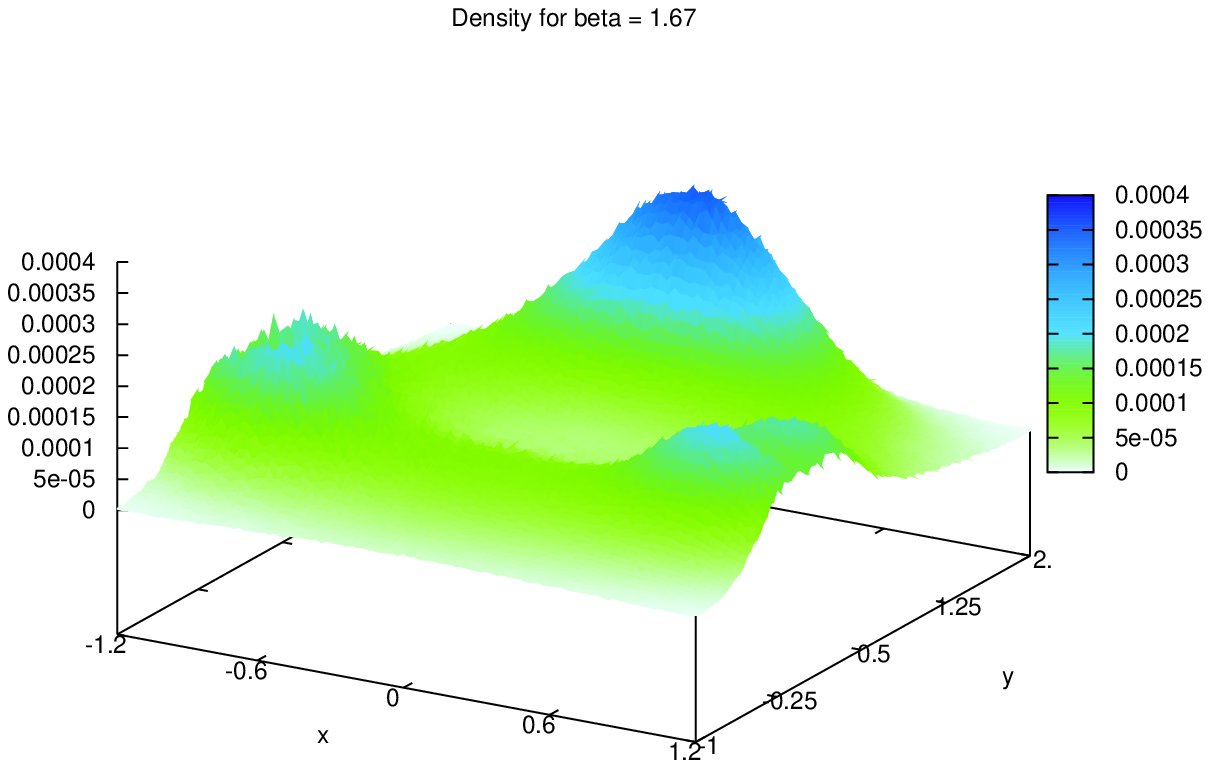,width=6cm}
 \epsfig{file=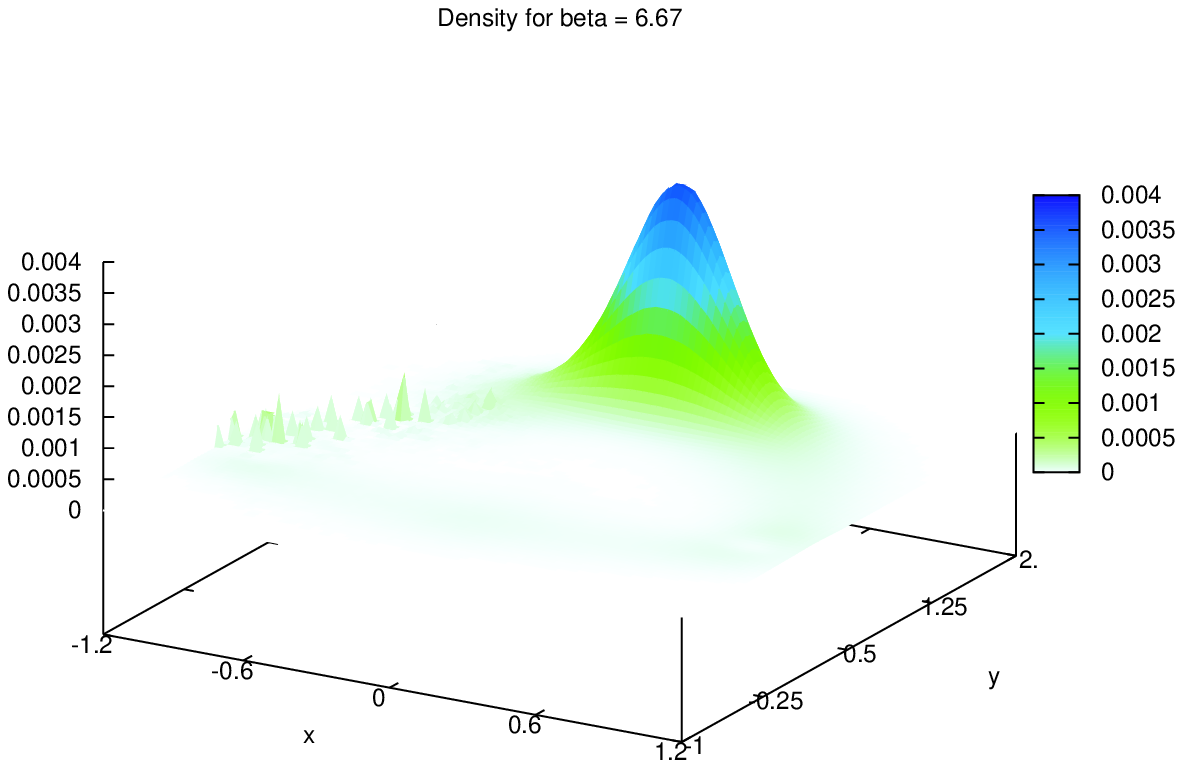,width=6cm}
 \epsfig{file=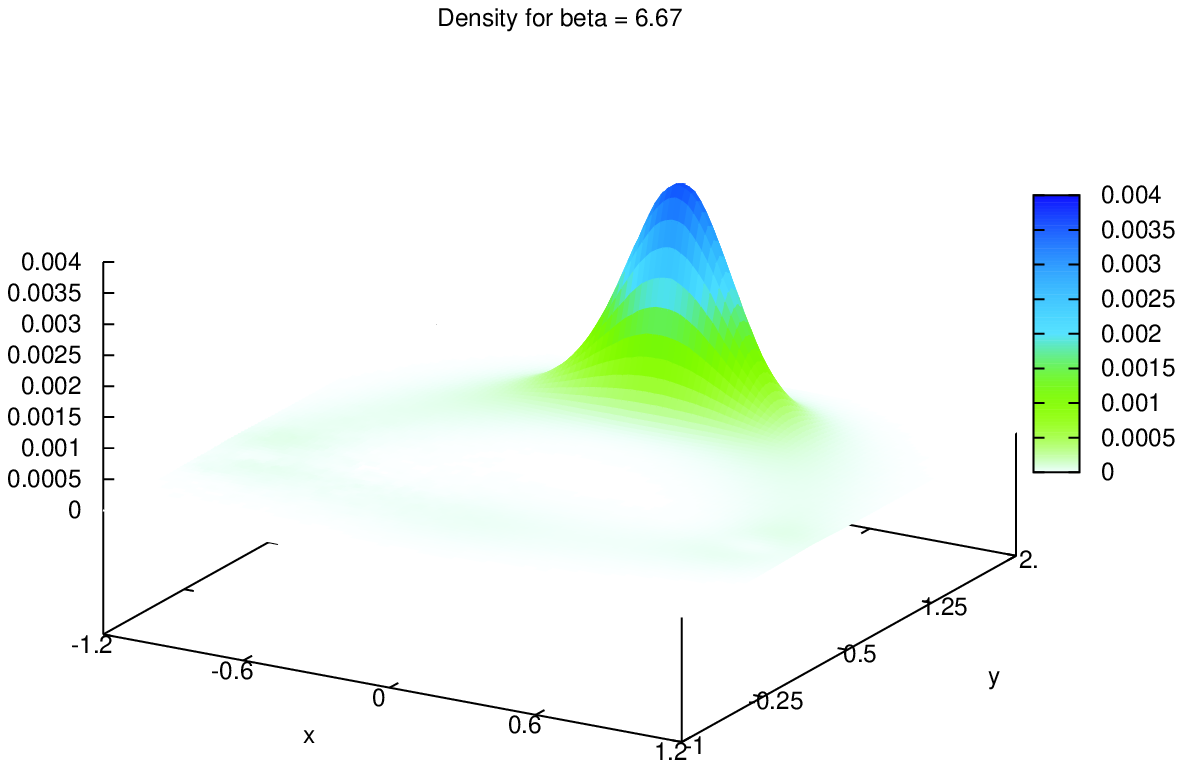,width=6cm}
\caption{Density $\rho$ for different choice of the reaction coordinate, of
  $\beta$, and of the number of paths $N$. The first column corresponds to the reaction coordinate $\xi_1$
  and the second to $\xi_2$. For the first two rows, the number of paths is
  $N=10^{4}$ and for the last two rows, $N=10^{5}$. The value for $\beta$ is given
  on each figure.}
  \label{fig:rho_coordinate1}
\end{figure}

Figure~\ref{fig:rho_coordinate1} gives the estimation of the density $\rho$ obtained with the two reaction coordinates~\eqref{eq:2RC}, for $N=10^{4},\, 10^5$ and for the two temperature values $\beta=1.67,\,6.67$. We observe that the numerical results are slightly better with $\xi_2$ (which is actually closer to the committor function $q$, see Figure~10 in~\cite{metzner-schuette-vanden-eijnden-06}): in particular, the selection procedure retains more diversity in the paths around~$A$ with this reaction coordinate. However, it can also be observed that the results obtained with these two reaction coordinates are all consistent: one does not need to know {\em a priori} the committor function to get reliable results. In the following, we concentrate on $\xi=\xi_2$, which is actually an interesting generic reaction coordinate since it only requires the knowledge of a reference configuration in $A$.

\begin{figure}
  \centering
 \epsfig{file=TGridd167_4.eps,width=6cm}
\epsfig{file=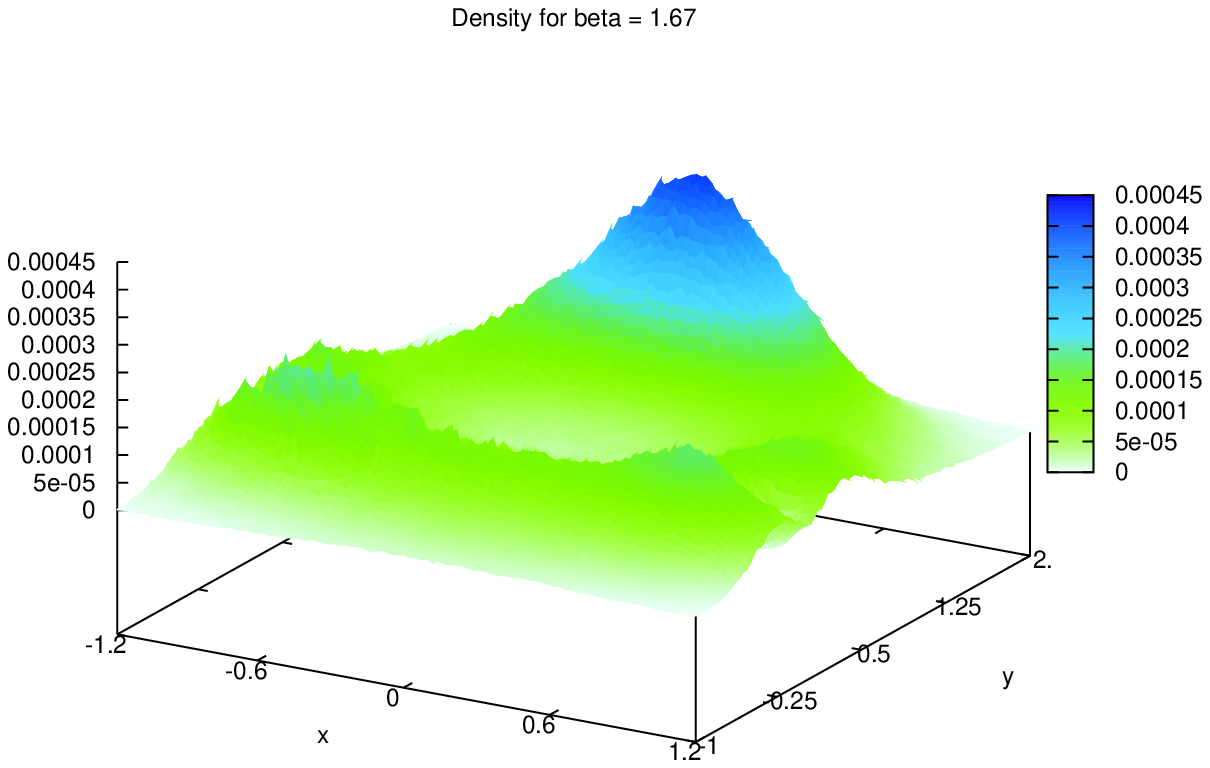,width=6cm}
 \epsfig{file=TGridd667_4.eps,width=6cm}  
 \epsfig{file=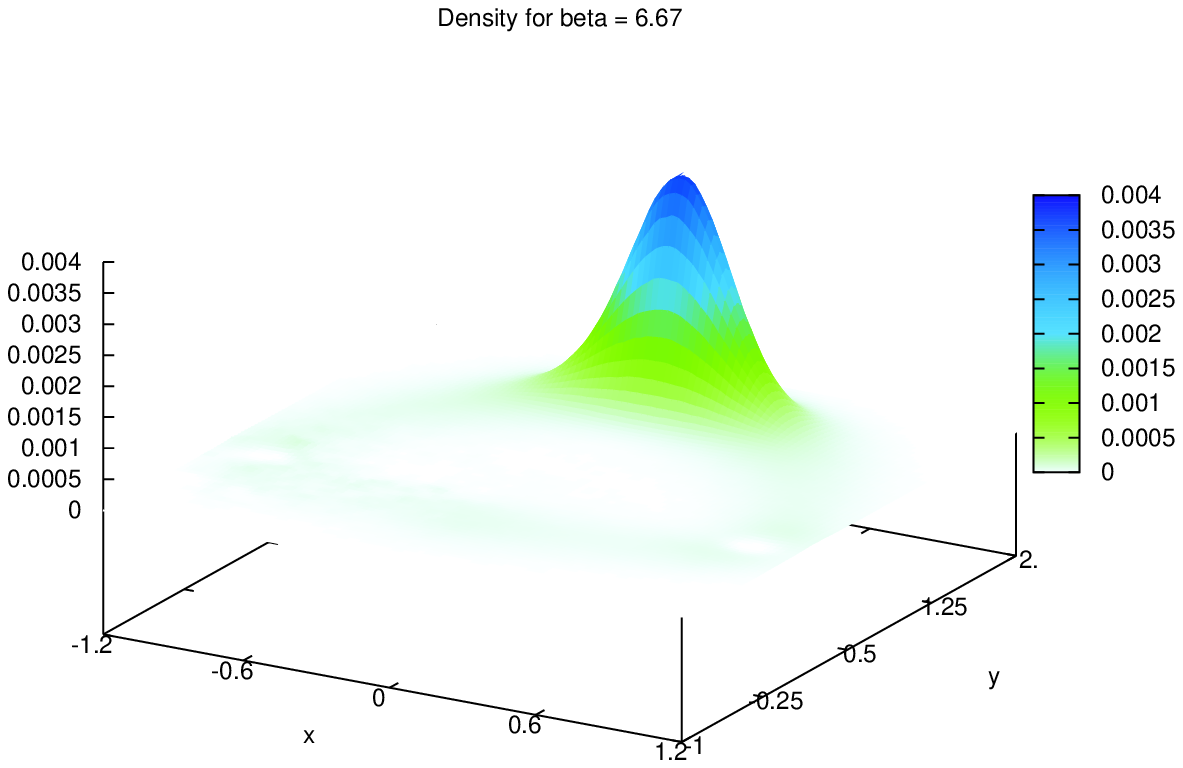,width=6cm}
  \caption{Density function $\rho$ for simulations with $\Delta t =
    10^{-2}$ (left column) and $\Delta t = 10^{-3}$ (right column). For this test
    $N=10^{4}$, the reaction coordinate is $\xi_2$ and the value of $\beta$ is given on each figure.}
  \label{fig:time_step_error}
\end{figure}

On Figure~\ref{fig:time_step_error}, we test the influence of the time-step size, by plotting the density $\rho$ obtained using two time-step sizes: $\Delta t= 10^{-2}$ (as above) and $\Delta t= 10^{-3}$. It appears that the results have indeed converged for the time-step size $\Delta t=10^{-2}$.

\begin{figure}[htbp]
  \centering
\epsfig{file=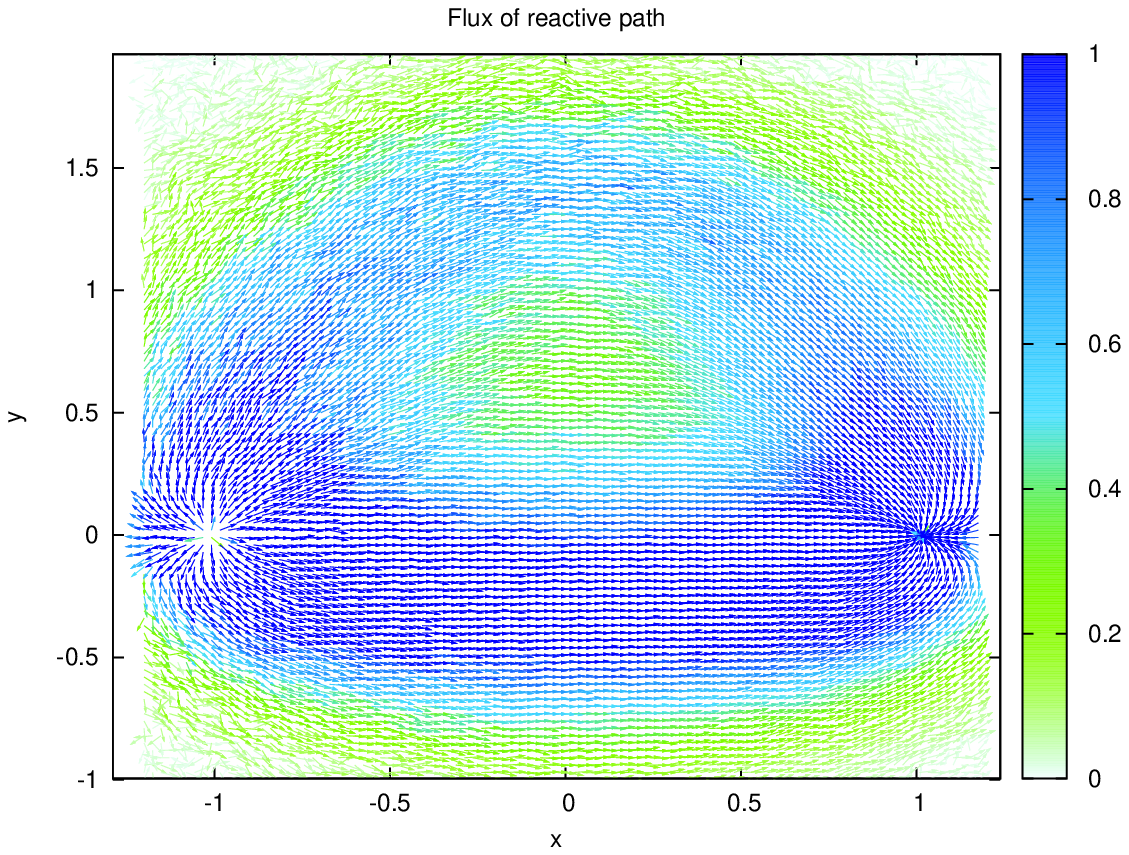,width=6.5cm}
\epsfig{file=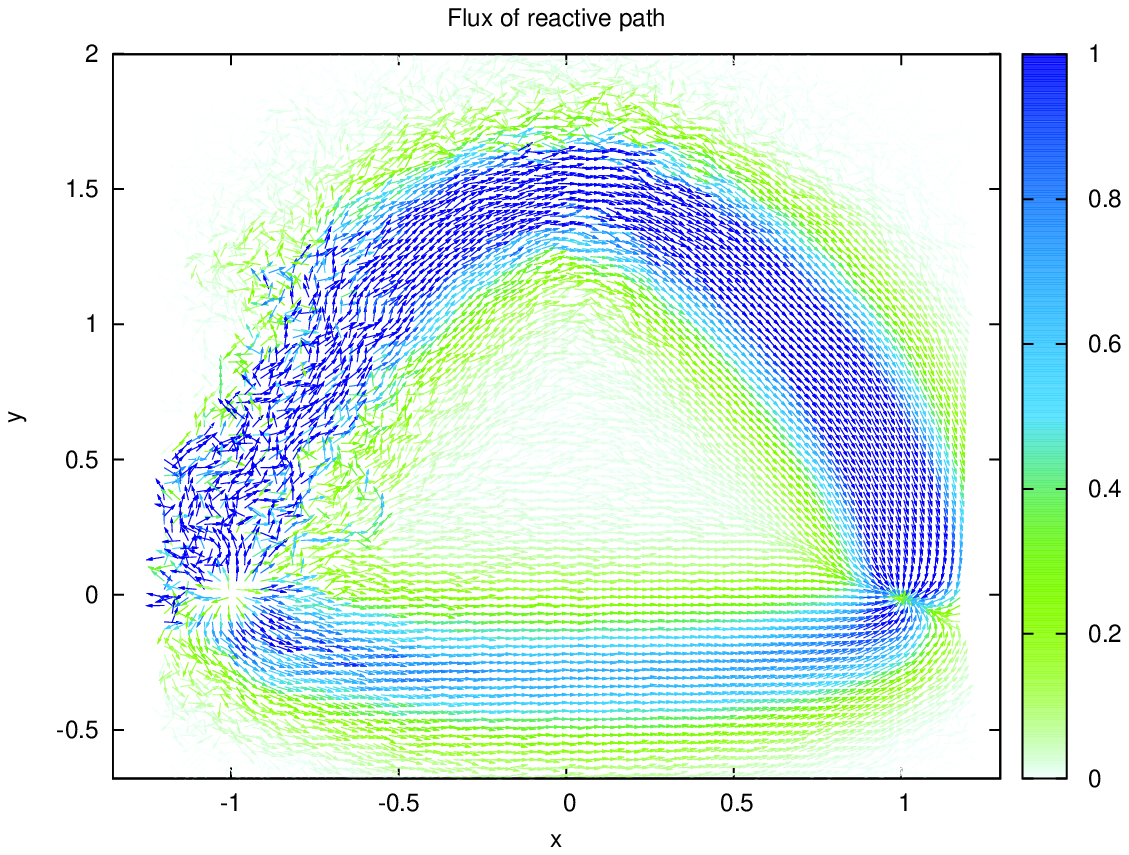,width=6.5cm}
\epsfig{file=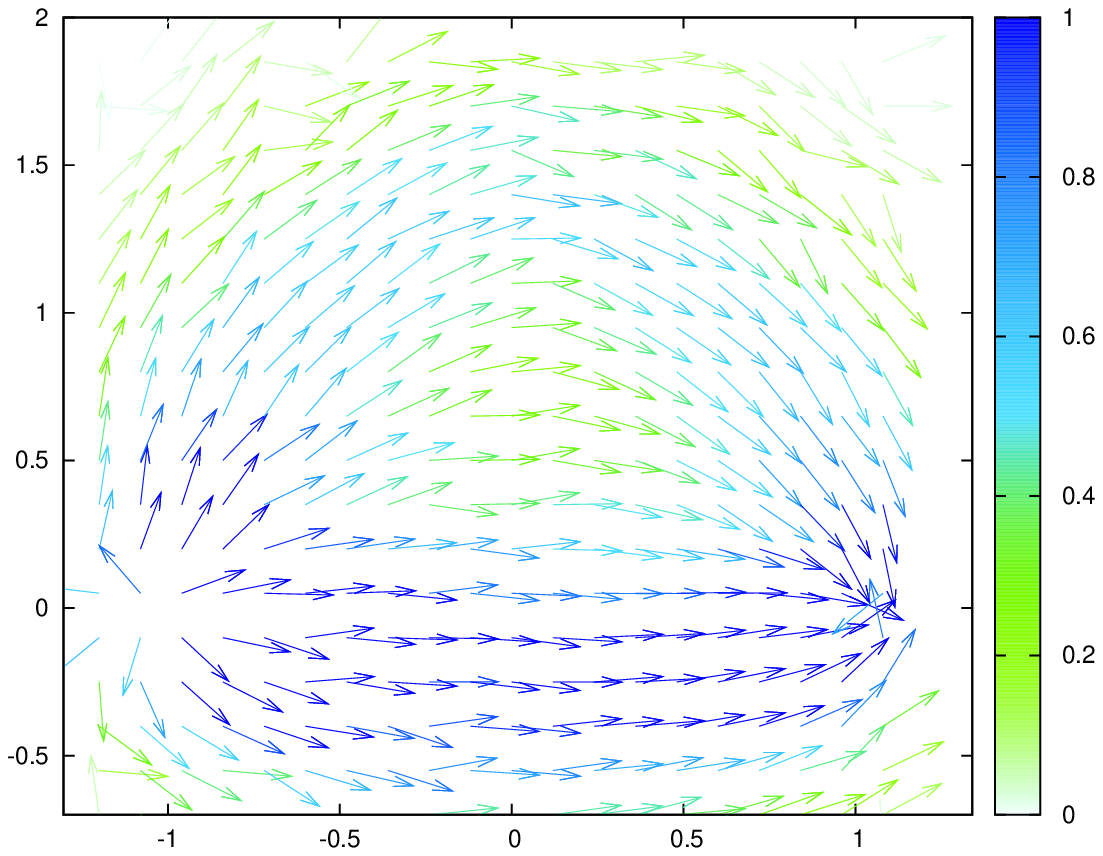,width=6.5cm}
\epsfig{file=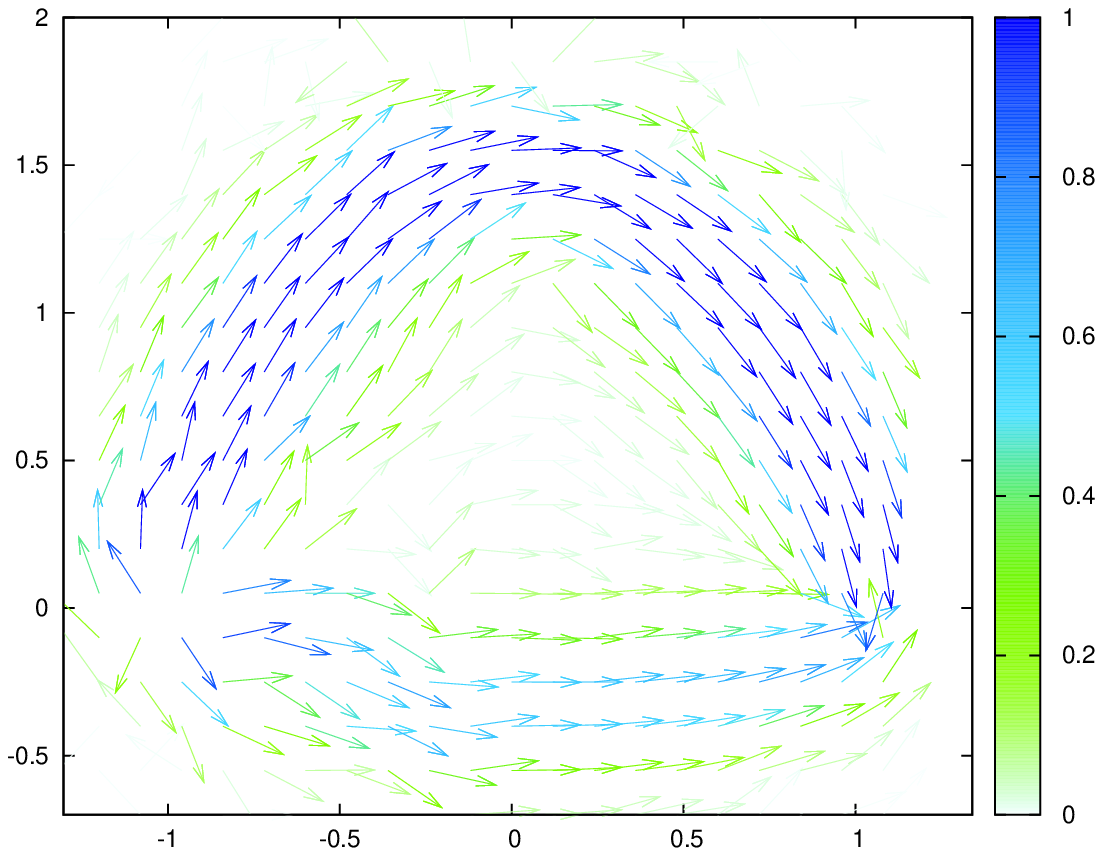,width=6.5cm}
\caption{Flux of reactive trajectories, at inverse temperature $\beta=1.67$ on the left, and $\beta=6.67$ on the right. The color indicates the norm of the flux.  }
  \label{fig:flux}
\end{figure}

An important quantity computed from reactive paths is the flux of reactive trajectories, which is defined (up to a multiplicative constant) as~\cite{hummer-04,metzner-schuette-vanden-eijnden-06}: for any domain ${\mathcal D} \in \R^d \setminus (\overline{A} \cup \overline{B})$,
$$\int_{\mathcal D} \div J= \P(\text{a reactive trajectory enters $\mathcal D$}) - \P(\text{a reactive trajectory leaves $\mathcal D$}).$$
On Figure~\ref{fig:flux}, we plot the flux $J$ at the two temperature
values. On the top figures, we use the grid of size $100 \times 100$. On the lower figures, we plot the flux on a sub-grid of
size $20 \times 20$ with constant $x$ and $y$ intervals to better illustrate
the overall direction. It is clear from these figures that at low temperature
($\beta = 6.67$), the upper channel is favored, while at higher temperature
($\beta =1.67$), the lower channel is more likely.

We would like to stress that all these results are in agreement with those obtained in~\cite{metzner-schuette-vanden-eijnden-06}, where similar results are obtained using a different numerical method.

\begin{figure}
  \centering
  \epsfig{file=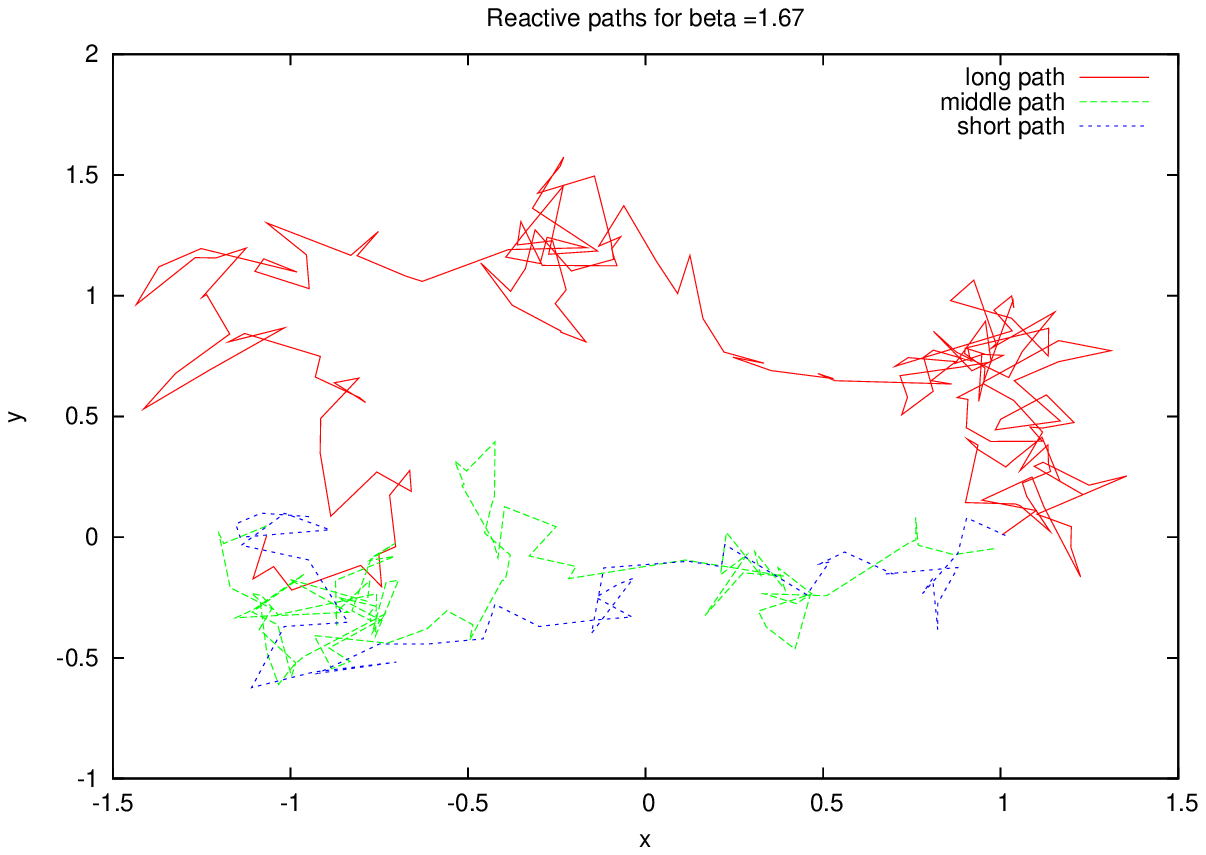,width=6cm}
  \epsfig{file=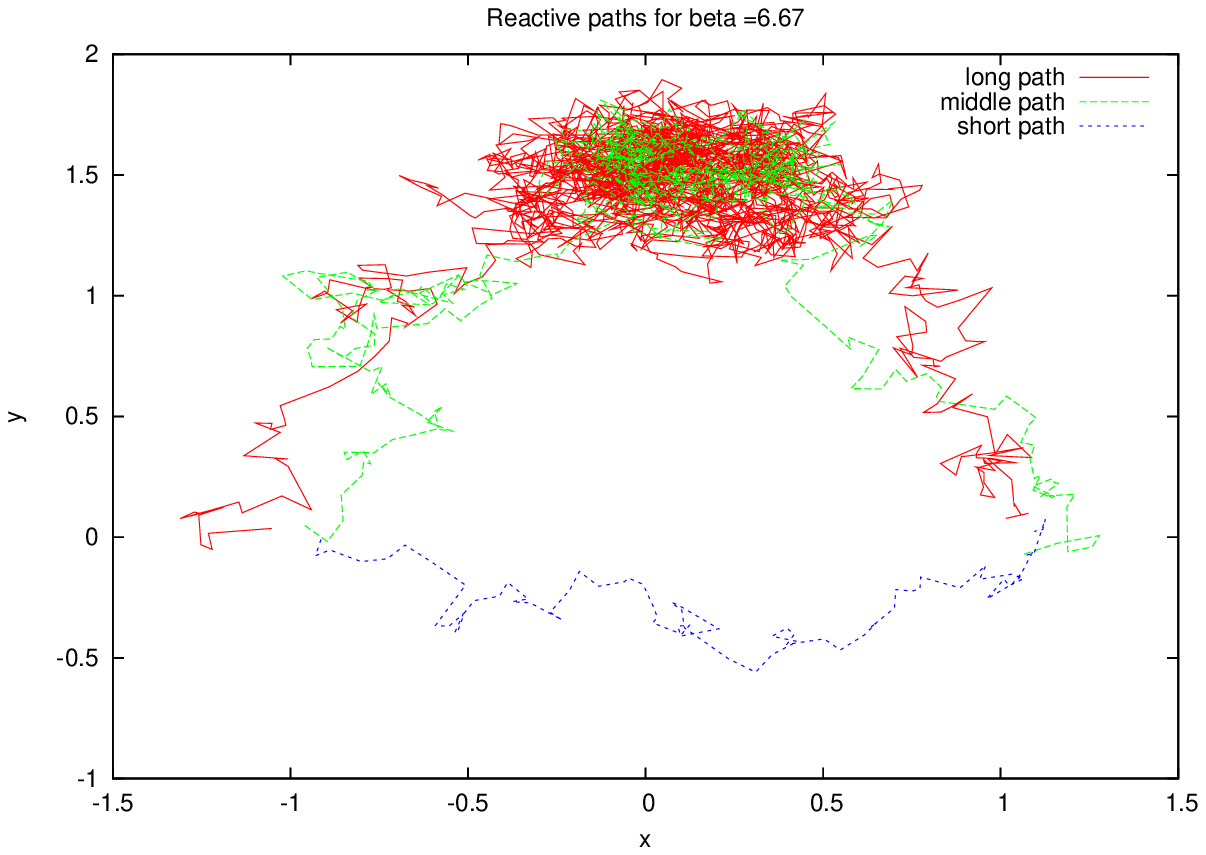,width=6cm}
 \caption{A few reactive paths for $\beta=1.67$ (left), $\beta=6.67$ (right).}
  \label{fig:reactive_paths_2d}
\end{figure}

On Figure~\ref{fig:reactive_paths_2d}, a few reactive paths are plotted. 
To quantify the fact that the upper or the lower channel is preferentially
used by reactive paths, let us consider $X^y_{\sigma_0}$ which is the $y$-value of
the reactive path at the first time $\sigma_0$ such that the $x$-value of the process $X_t$ is
equal to $0$. We consider that the reactive path goes through the upper (resp.
the lower) channel if $X^y_{\sigma_0}$ is larger than
$0.75$ (resp. smaller than $0.25$).
For $\beta=6.67$, the proportion of paths such that $0.25 \leq X_{\sigma_0}^y \leq
0.75$ is  $0.28\%$ and the paths going through the upper (resp. the
lower) channel  is $62.55\%$ (resp. $37.17 \%$). 
For $\beta=1.67$, the proportion of paths such that $0.25 \leq X_{\sigma_0}^y \leq
0.75$ is  $11.26\%$ and the paths going through the upper (resp. the
lower) channel  is $31.46\%$ (resp. $57.28 \%$).

\begin{figure}
  \centering
  \epsfig{file=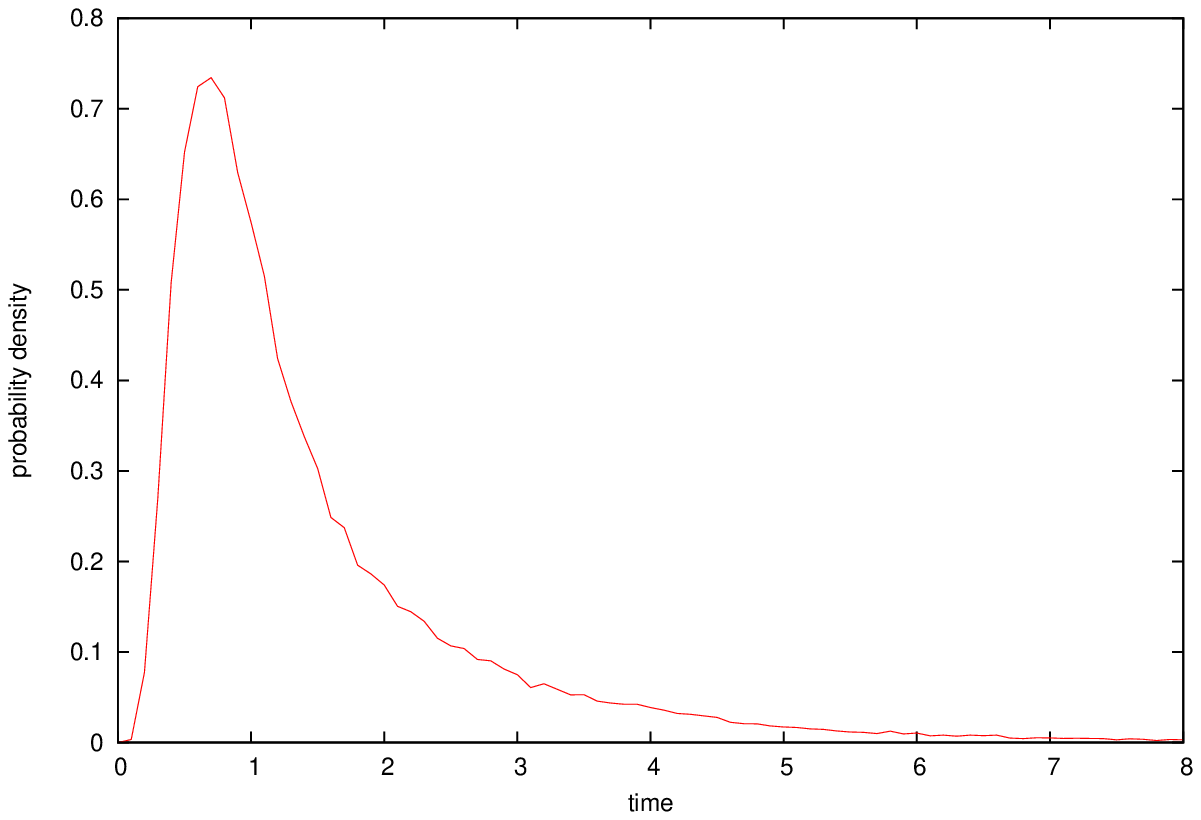,width=6cm}
  \epsfig{file=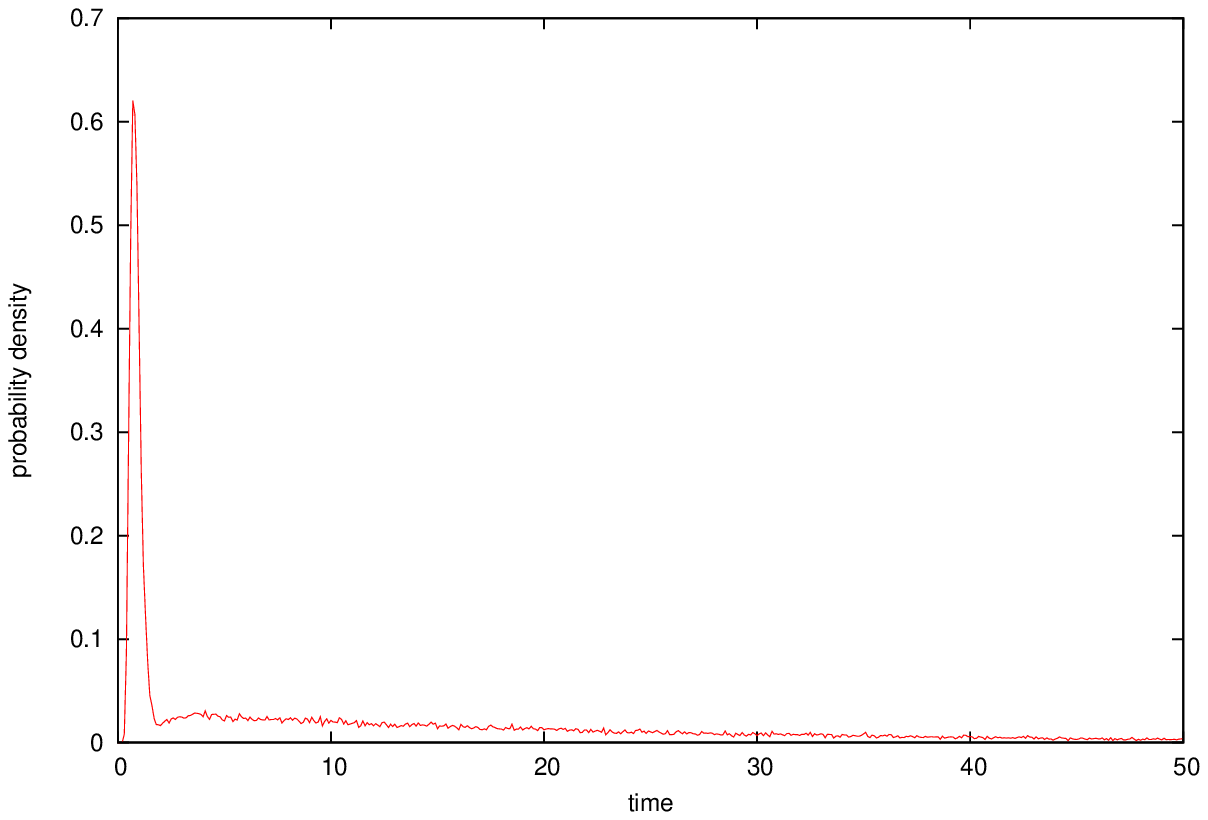,width=6cm}
  \epsfig{file=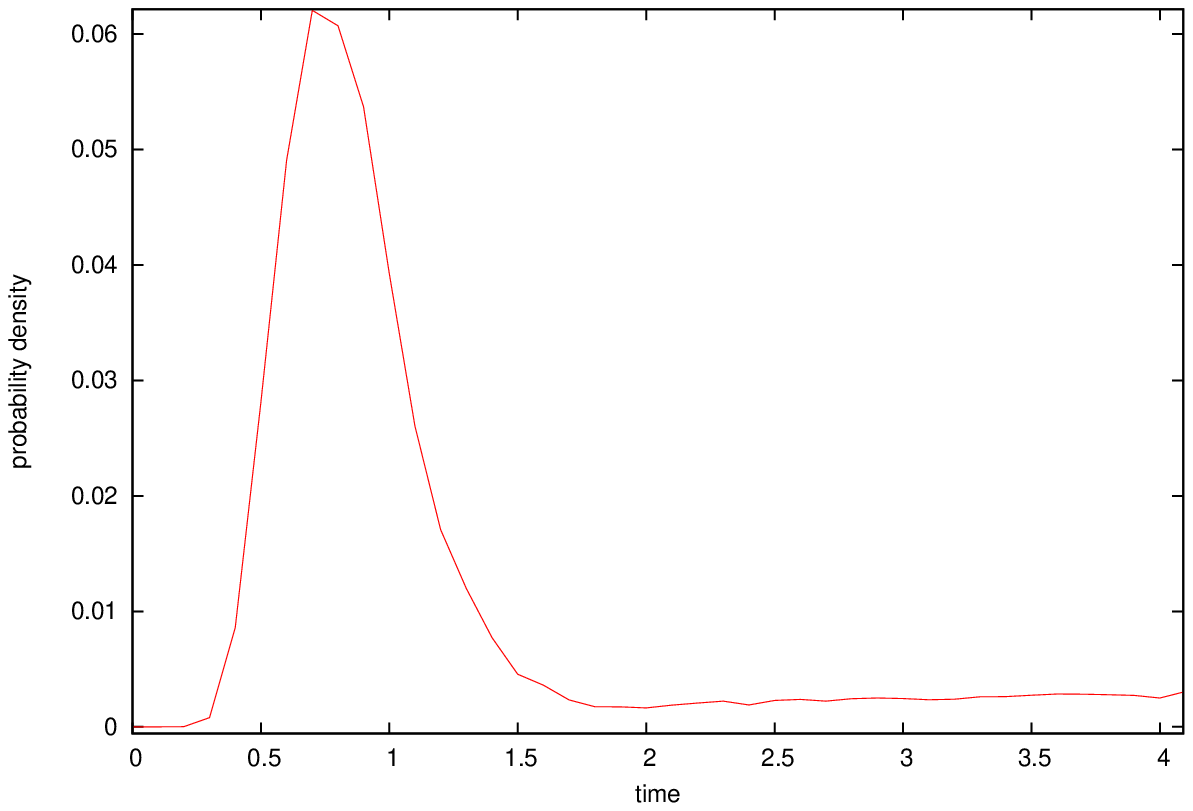,width=6cm}
  \epsfig{file=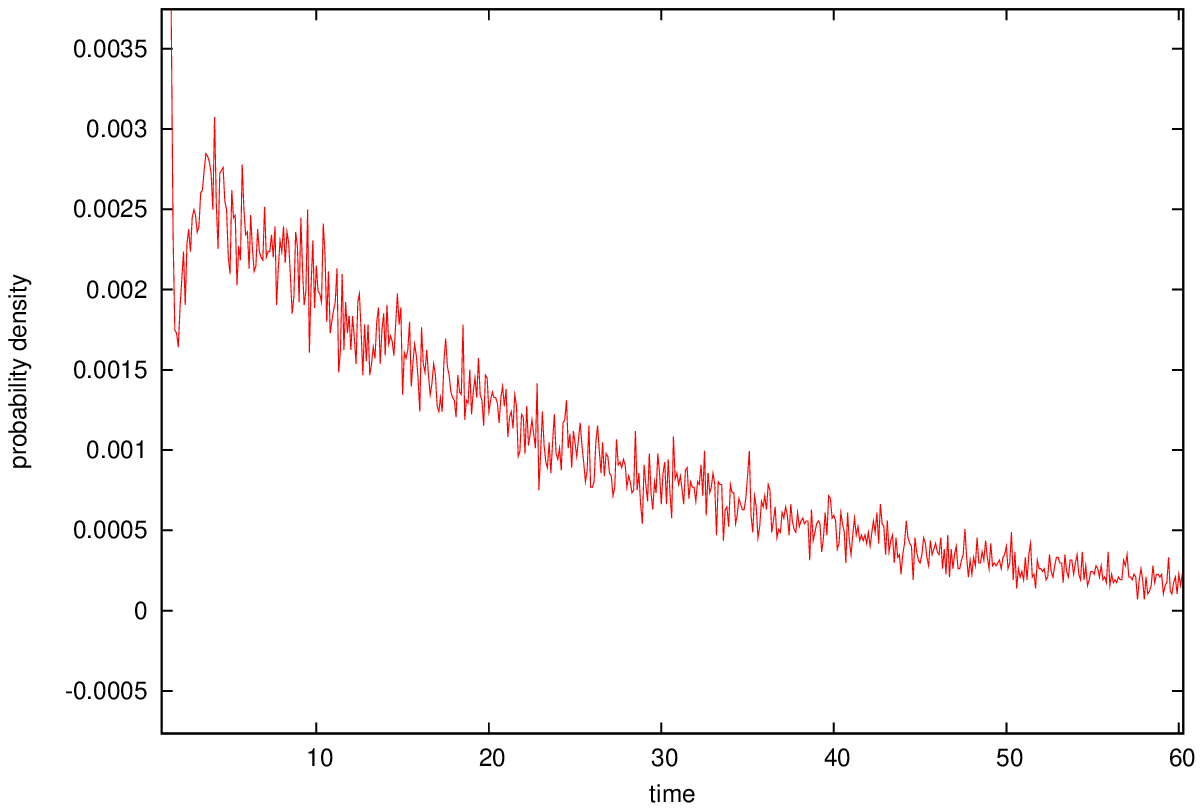,width=6cm}
 \caption{Distribution of the time lengths of reactive paths for 
    $\beta=1.67$ (top left), $\beta=6.67$ (top right) and zoom on the two modes of the
    distribution for $\beta=6.67$ (bottom). The distribution of the bottom
    left (resp. bottom right) corresponds to the time lengths of paths through the lower channel
    (resp the upper channel).}
 \label{fig:distribution_time_2d}
\end{figure}

Finally, we plot on Figure~\ref{fig:distribution_time_2d}, the histogram of
the time lengths of reactive trajectories, at the two temperatures. We observe
two modes in this distribution when $\beta=6.67$, corresponding to the two
channels. These two modes overlap when $\beta=1.67$.

\section{Computing transition times}
\label{sec:4}
In this section, we would like to explain how to use the algorithm above to compute transition times, namely the time required to go from one metastable state to another one. In the notation introduced above, it is typically the time required to go from $\partial A$ to $\partial B$. A trajectory starting from $\partial A$ (with a distribution to be made precise) and reaching $\partial B$ is called a transition path. It is different from a reactive path: a reactive path is only the last portion of a transition path, linking $\partial A$ to $\partial B$ without going back to $A$.

The procedure proposed in this section will be tested numerically in Section~\ref{sec:nd_T} on the two cases considered above: the double-well potential of Section~\ref{sec:1d} and the two dimensional potential of Section~\ref{sec:2d}.

\subsection{The one-dimensional case}

Let us first explain the principle of the algorithm in the one-dimensional setting, for simplicity. In this case, $A$ and $B$ are singletons, say $A=\{-1\}$ and $B=\{+1\}$ and the reaction coordinate is simply $\xi(x)=x$ so that $\Sigma_{z}=\{z\}$. One could think of the double-well potential considered in Section~\ref{sec:1d}.

The time we are interesting in here is the time needed to go from $A$ to $B$, namely
\begin{equation}\label{eq:T1d}
T=\inf \{t>0, X_t \in B \} 
\end{equation}
where $X_t$ solves~\eqref{eq:sde} with initial condition
$$X_0 = -1.$$
Note that this time $T$ is typically much larger than the time length of a reactive trajectory $\tau$ considered above.

To compute this time $T$, let us consider $\Sigma_{z_{\rm min}} = \{z_{\rm min}\}$ and cut a trajectory which ends in $B$ into several pieces. Starting from $A$, the trajectory will touch $\Sigma_{z_{\rm min}}$, within a time $T^1_1$. Then, two events may occur. Either the trajectory goes to $B$ before reaching $A$ (with probability $p$, and within a time $T_3$) or it goes back to $A$ before reaching $B$ (with probability $(1-p)$, and within a time $T_2^1$). By iterating the argument and using the Markov property, one can check that:
\begin{equation}\label{eq:T_formula}
T = \sum_{i=1}^{I-1} (T_1^i + T_2^i) + (T_1^I + T_3)
\end{equation}
where, for $i \ge 1$, $T_1^i$ (resp. $T_2^i$) are identically distributed random variables, and~$I$ is a geometric random variable independent from $(T_1^i,T_2^i,T_3)$ with parameter~$p$: for $i \in \{1,2, \ldots \}$, $\P(I=i)=(1-p)^{i-1}p$. In~\eqref{eq:T_formula}, $\sum_{i=1}^0 \cdot =0$ by convention.

From~\eqref{eq:T_formula} and using the fact that $\E(I)=1/p$, it is easy to check that
\begin{equation}\label{eq:esp_T_formula}
\E(T) = \left(\frac{1}{p} - 1\right) \E(T_1 + T_2) + \E(T_1 + T_3),
\end{equation}
where we denoted $T_1=T_1^1$ the time for a trajectory starting at $-1$ to go to $z_{\rm min}$ and $T_2=T_2^1$ the time for a trajectory starting from $z_{\rm min}$ to go back to $-1$ without touching $1$.

Let us now explain how we evaluate numerically each term which appears in~\eqref{eq:esp_T_formula}:
\begin{itemize}
\item The average time $\E(T_1 + T_2)$ which corresponds to the mean time for a trajectory starting from $A$, to reach $\Sigma_{z_{\rm min}}$ and then to go back to $A$ without touching $B$ is obtained by DNS. Indeed, since $\Sigma_{z_{\rm min}}$ is typically close to $A$, this is easily obtained by simple Monte Carlo simulations.
\item The average time $\E(T_1 + T_3)$ which corresponds to the mean time for a trajectory starting from $A$, to reach $\Sigma_{z_{\rm min}}$ and then to go to $B$ without touching again $A$ is obtained by a slight modification of the algorithm of Section~\ref{sec:algo_details}. Namely, in Step 2 of the initialization procedure, we do not discard the beginning of the trajectory to get the whole time, starting from $A$, to reach $\Sigma_{z_{\rm min}}$ (taking into account all the recrossings of $\partial A$).
\item Finally, the probability $p$ which is the probability for a trajectory starting from $\Sigma_{z_{\rm min}}$ to reach $B$ before touching $A$ is also obtained by a slight modification of the algorithm of Section~\ref{sec:algo_details}. Namely, the initialization steps (1 to 3) are simply replaced by setting $X_0=z_{\rm min}$. Then, $\hat{\alpha}_N$ defined by~\eqref{eq:alphaN} yields an estimate of $p$.
\end{itemize}
Let us now explain how to generalize this approach to high-dimensional situations.

\subsection{The general case}
\label{sec:trans_time_gen}

We would like to generalize the method above to higher dimension. In such a case and compared to the previous section, $\partial A$ (resp. $\partial B$) plays the role of $-1$ (resp $1$). Again, a transition path is cut into pieces: a first part of duration $T^1_1$ between $\partial A$ and $\Sigma_{z_{\rm min}}$, then (possibly) a first part of duration $T^1_2$ between $\Sigma_{z_{\rm min}}$ back to $\partial A$, ... until a $I$-th piece of duration $T^I_1$ between $\partial A$ and $\Sigma_{z_{\rm min}}$ and a final piece of duration $T_3$ between $\Sigma_{z_{\rm min}}$ and $\partial B$.

The difficulty when generalizing the reasoning above is to keep the Markovian properties which have been used to cut the trajectories into pieces and obtain formula~\eqref{eq:esp_T_formula}, namely:
\begin{itemize}
\item[(i)] The fact that the number $I$ of trajectories from $\partial A$ to $\Sigma_{z_{\rm min}}$ before observing the transition to $\partial B$ is a geometric variable with parameter $p$, {\em independent of the times $(T_1^i,T_2^i,T_3)$}. This requires that for a trajectory coming from $\partial A$ and reaching $\Sigma_{z_{\rm min}}$, the probability of reaching $B$ before $A$ does not depend on the hitting point on $\Sigma_{z_{\rm min}}$.
\item[(ii)] The fact that the distribution $\nu_{\partial A}$ of starting
  points on $\partial A$ and the associated distribution $\nu_{\Sigma_{z_{\rm
        min}}}$ of first hitting points on $\Sigma_{z_{\rm min}}$ are
  ``balanced'' in the sense that: starting from $\nu_{\partial A}$ on
  $\partial A$, the first hitting points on $\Sigma_{z_{\rm min}}$ are
  distributed according to $\nu_{\Sigma_{z_{\rm min}}}$ ; and starting from
  $\nu_{\Sigma_{z_{\rm min}}}$ on $\Sigma_{z_{\rm min}}$, the distribution of
  first hitting points on $\partial A$ is again $\nu_{\partial A}$. 
% \comment{conditionellement au fait qu'on ne touche pas $B$ ?}
% Oui mais ce n'est pas grave: tu regardes le premier temps ou tu touches $A$ ou $B$ an partant de $\Sigma_{z_{\rm min}}$ et tu as sur 
% $\partial A$ la bonne distrib.
\end{itemize}
These two conditions are satisfied if specific $\Sigma_{z_{\rm min}}$, $\nu_{\partial A}$  and $\nu_{\Sigma_{z_{\rm min}}}$ are considered (see~\cite{vanden-eijnden-venturoli-ciccotti-elber-08} for related considerations):
\begin{proposition}
Let us consider $X_t$ solution to~\eqref{eq:sde} and the associated committor function $q$ defined by~\eqref{eq:committor}. Let us assume that
$$\text{(H1) $\Sigma_{z_{\rm min}}$ is an isocommittor surface}$$
namely that there exists a $q_0 \in (0,1)$ such that
$$\Sigma_{z_{\rm min}}=\{ x \in \R^d, \, q(x)=q_0\}.$$
Then, property (i) above is satisfied.
Assume moreover that:
$$\text{(H2) $X_0$ is distributed according to $d\nu_{\partial A}= Z_{\partial A}^{-1} |\nabla q| e^{-\beta V} d \sigma_{\partial A}$.}$$
Then, the distribution of first hitting times on $\Sigma_{z_{\rm min}}$ is $$d\nu_{\Sigma_{z_{\rm min}}}= Z_{\Sigma_{z_{\rm min}}}^{-1} |\nabla q| e^{-\beta V} d \sigma_{\Sigma_{z_{\rm min}}},$$
and property (ii) above is satisfied.
\end{proposition}
In this proposition, the measures $\sigma_{\partial A}$ and $\sigma_{\Sigma_{z_{\rm min}}}$ are the Lebesgue measures on $\partial A$ and $\Sigma_{z_{\rm min}}$ induced by the Lebesgue measure in the ambient space $\R^d$, and the Euclidean scalar product.
\begin{proof}
Let us first start with property (i), assuming (H1). By definition of the committor function $q$ (see~\eqref{eq:committor_interpretation}), for any point $x \in \Sigma_{z_{\rm min}}$, the probability to reach~$B$ before~$A$ is~$p$ (independently of $x$), and thus, the number of trajectories from $\partial A$ to $\Sigma_{z_{\rm min}}$ before a transition to $\partial B$ is a geometric random variable with parameter $p$, independent on the times $(T_1^i,T_2^i,T_3)$ introduced above.

For property (ii), let us consider a diffusion $X_t$ solution to~\eqref{eq:sde} with initial conditions $X_0$ distributed according to $\nu_{\partial A}$. The aim is to prove that, for any fixed test function $\varphi: \R^d \to \R$,
\begin{equation}\label{eq:proof_nu}
\E(\varphi (X_{T_1}))= \int_{\Sigma_{z_{\rm min}}} \varphi \, d\nu_{\Sigma_{z_{\rm min}}},
\end{equation}
where $T_1= \inf \{t \ge 0, X_t \in \Sigma_{z_{\rm min}} \}$. Let us denote
$$f(x)=\E(\varphi (X^x_{T_1^x}) )$$
where $X_t^x$ is the solution to~\eqref{eq:sde} with starting point $X_0=x$, and $T_1^x$ the associated first hitting time of $\Sigma_{z_{\rm min}}$. Then,~\eqref{eq:proof_nu} amounts to proving that:
$$ Z_{\partial A}^{-1} \int f \exp(-\beta V) |\nabla q| \, d \sigma_{\partial A} = Z_{\Sigma_{z_{\rm min}}}^{-1} \int f \exp(- \beta V) |\nabla q| \, d \sigma_{\Sigma_{z_{\rm min}}},$$
or equivalently
\begin{equation}\label{eq:proof_nu_dirac}
 Z_{\partial A}^{-1} \int f \exp(-\beta V) |\nabla q|^2  \, \delta_{q(x)}(dx) = Z_{\Sigma_{z_{\rm min}}}^{-1} \int f \exp(- \beta V) |\nabla q|^2 \, \delta_{q(x)-q_0}(dx),
\end{equation}
where we used the measure $\delta_{q(x)-r}(dx)$ defined by:
$$\delta_{q(x)-r}(dx) = |\nabla q|^{-1} \, d \sigma_{\{x, q(x) = r \}}.$$
This measure may also be seen (using the co-area formula, see for example~\cite[Lemma 3.2]{lelievre-rousset-stoltz-10}) as a conditional measure such that: for any test function $\psi$
$$\int \psi(x) \, dx = \int \int_{\{x, q(x) = r \}} \psi \, \delta_{q(x)-r}(dx) \, dr.$$
One can check the following important derivation property (see for example~\cite[Lemma 3.10]{lelievre-rousset-stoltz-10}):
\begin{equation}\label{eq:deriv_dirac}
\frac{d}{dr} \int_{\{x, q(x) = r \}} \psi \, \delta_{q(x)-r}(dx) = \int_{\{x, q(x) = r \}} \div( \psi \nabla q |\nabla q|^{-2} ) \, \delta_{q(x)-r}(dx).
\end{equation}
Let us go back to the proof of~\eqref{eq:proof_nu_dirac}. What can actually be checked is that $I'(r)=0$ where
$$I(r)=\int f \exp(-\beta V) |\nabla q|^2 \,  \delta_{q(x)-r}(dx),$$
which implies~\eqref{eq:proof_nu_dirac} (consider the case $\varphi= f = 1$ to get that  $ Z_{\partial A} =  Z_{\Sigma_{z_{\rm min}}}$).
Indeed, we have (using~\eqref{eq:deriv_dirac}):
\begin{align*}
I'(r)
&= \int \div( f \exp(-\beta V) \nabla q) \, \delta_{q(x)-r}(dx)\\
&= \int \nabla f \cdot \nabla q \exp(-\beta V) \, \delta_{q(x)-r}(dx) +  \int \div(\exp(-\beta V) \nabla q)  f \, \delta_{q(x)-r}(dx).
\end{align*}
The second term is zero using~\eqref{eq:committor}. Likewise, the first term is zero since, for any test function $h: \R \to \R$,
\begin{align*}
\int h(r) \int \nabla f \cdot \nabla q \exp(-\beta V) \, \delta_{q(x)-r}(dx) \, dr
&= \int h \circ q  \, \nabla f \cdot \nabla q \exp(-\beta V) \\
&= \int   \nabla f \cdot \nabla (H \circ q ) \exp(-\beta V) \\
&= - \int \div ( \nabla f \exp(-\beta V) ) H \circ q \\
& = 0
\end{align*}
where $\circ$ denotes the composition operator, and $H$ is a primitive of $h$. This concludes the proof.
\end{proof}
Thus, under assumptions (H1) and (H2), formula~\eqref{eq:esp_T_formula} above holds with~$T$ being defined as the transition time from $\partial A$ to $\partial B$ and $X_0$ being distributed according to $\nu_{\partial A}$. Even though this measure may not appear to be very natural, the result should not be very sensitive to the choice of the distribution on $\partial A$ since $A$ is assumed to be a metastable state so that any trajectory starting from $\partial A$ will rapidly loose the memory of its starting point after a mixing time within $A$. One simple interpretation of this measure is that it is the stationary measure of the Markov chain which, to one point of $\partial A$ associates the next point on $\partial A$ for the dynamics~\eqref{eq:sde} after $\Sigma_{z_{\rm min}}$ has been touched. Note eventually that the techniques presented in Section~\ref{sec:nd_T} to evaluate $\E(T_1 + T_2)$, $\E(T_1 + T_3)$ and $p$ can also be applied in higher dimensional cases.

A few remarks are in order. First concerning (H1), it is in general impossible to find a precise approximation of an isocommittor in high dimension. However, the algorithm we propose only requires such an isocommittor {\em in a neighborhood of $A$} and thus, a rough approximation may often be obtained by simple considerations, like defining $\Sigma_{z_{\rm min}}$ as the configurations at a given (small) distance of~$A$. We will check below on a two-dimensional case that a rough
approximation $\Sigma_{z_{\rm min}}$ of an isocommittor in a
neighborhood of $A$ does not imply a too large error on the estimated
mean transition time. Of course, this remains to be investigated theoretically, or numerically for more realistic test cases. It would be interesting to show, using the metastable features of $A$, that the time lengths of the cycles from $A$ to $A$ are actually close to be independent. This can be done for Markov chains introducing so-called regeneration times (see for example~\cite{sigman-wolff-93}). We are currently investigating if this approach could yield a justification of~(\ref{eq:T_formula}) without assumption (H1).
Second, concerning (H2), a simple way to generate an ensemble of configurations distributed according to $\nu_{\partial A}$ is to subsample the Markov chains of visiting points to $\partial A$ for excursions leaving $\partial A$, touching $\Sigma_{z_{\rm min}}$, an then going back to $\partial A$ (as mentioned above). This initialization procedure could also be used alternatively to steps 1-2-3 in the algorithm described in Section~\ref{sec:algo_details}.

\section{Computing transition times: numerical illustrations}
\label{sec:nd_T}
\subsection{The one-dimensional case}\label{sec:1d_T}

Let us consider again the one-dimensional double-well potential already used above in Section~\ref{sec:1d}, with $A=\{-1\}$, $B=\{1\}$ and $\xi(x)=x$. Recall that the transition time is defined as
$$T=\inf \{t > 0, X_t=1 \, \vert \, X_0 =-1 \}.$$
Using the notation of Section~\ref{sec:4}, 
let us also define
$$T_1=\inf \{t > 0, X_t= z_{\rm min} \, \vert \, X_0 = -1 \},$$
$$T_2=\inf \{t > 0, X_t =-1 \, \vert \, X_0 = z_{\rm min} \text{ and
}\forall s \in (0,t), \  X_s \neq 1 \},$$
$$T_3=\inf \{t > 0, X_t =1   \, \vert \, X_0 = z_{\rm min}\text{ and }
\forall s \in (0,t), \  X_s \neq -1 \}.$$
We use formula~\eqref{eq:esp_T_formula} to estimate $\E(T)$.

First, in Tables~\ref{tab:meanoft1t2} and~\ref{tab:compa_transition_proba_cerou_dns}, we compare results obtained from DNS and from our algorithm using~\eqref{eq:esp_T_formula}, for small values of $\beta$ ($\beta \le 7$). In Table~\ref{tab:meanoft1t2}, we give the results of the transition time estimation for different values of
$\beta$, $\Delta t$ and~$z_{\rm min}$. Our estimation converges
to the DNS estimation when $\Delta t$ decreases. The difference between the results obtained by DNS and by our algorithm is smaller when $z_{\rm min}$ increases or $\beta$
increases. For $\beta=7$, we obtain an error of less than $0.5 \%$ compared to
a DNS. It seems that the main source of error is due to an underestimation of the
probability to hit back $A$ when starting from $\Sigma_{z_{\rm min}}$. This is related to the
simple Euler discretization of the diffusion~\eqref{eq:sde}. This fact is
illustrated on Table~\ref{tab:compa_transition_proba_cerou_dns}, where we see
that the error on $p$ decreases when $z_{\rm {min}}$ increases, $\Delta t$ decreases or $\beta$
increases.

Notice that the confidence intervals (C.I.) given in the tables are $95\%$ confidence intervals, estimated with $10$ independent runs.

\begin{figure}[htbp]
  \centering
  \epsfig{file=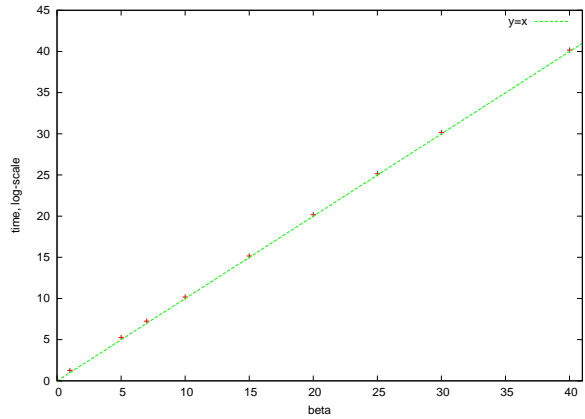,width=8cm}
  \caption{Comparison of the estimated mean transition times as a function of $\beta$ with the asymptotic law from large deviation theory.}
  \label{fig:asymptotic_time}
\end{figure}

Finally, results for larger values of $\beta$ are given in  Table~\ref{tab:1d_res_proba_time}. Note that $\E(T_3)$ is not significant in the estimation of $\E(T)$ for large values of $\beta$. We also check (see Figure~\ref{fig:asymptotic_time}) that
the numerical results satisfy the theoretical asymptotic behaviour obtained from the large deviation theory:
\begin{equation}
  \label{eq:11}
\E(T) \varpropto \exp (\beta \Delta V),
\end{equation}
where $\Delta V = 1$ is the height of the energy barrier between $-1$ and $1$.

\begin{table}
   \centering
  \begin{tabular}{|c|c|c|c|c|c|c|} \hline
    $\beta	$&$\Delta t	$&$z_{\rm min}$&$\E(T)$& $\E(T)$
    &C.I. on $\E(T)$ 	&Error	\\ 
    & & &(DNS)& (algo)   &(algo)	& ($\%$)	\\ \hline
    $1	$&$0.010	$&$-0.9	$&$3.634	$&$4.331	$&$[4.214,4.452]	$&$19.170	$\\
    $1	$&$0.010	$&$-0.8	$&$3.634	$&$4.149	$&$[4.055,4.247]	$&$14.185	$\\
    $1	$&$0.010	$&$-0.6	$&$3.634	$&$3.884	$&$[3.818,3.952]	$&$6.892	$\\
    $1	$&$0.001	$&$-0.9	$&$3.634	$&$3.911	$&$[3.798,4.029]	$&$7.620	$\\
    $1	$&$0.001	$&$-0.8	$&$3.634	$&$3.734	$&$[3.646,3.825]	$&$2.759	$\\
    $1	$&$0.001	$&$-0.6	$&$3.634	$&$3.533	$&$[3.466,3.602]	$&$2.775	$\\
    $5	$&$0.010	$&$-0.9	$&$185	   $&$208.3$&$[199.6,217.7]	$&$12.591	$\\
    $5	$&$0.010	$&$-0.6	$&$185	   $&$221.2$&$[214.3,228.4]	$&$19.577	$\\
    $5	$&$0.001	$&$-0.9	$&$185	   $&$187.4$&$[180.5,194.8]  $ &$1.292	$\\
    $5	$&$0.001	$&$-0.6	$&$185	   $&$193.2$&$[188.3,198.3]	$&$4.459	$\\
    $7	$&$0.010	$&$-0.9	$&$1405	  $&$1515$&$[1468,1565]	$&$7.832	$\\
    $7	$&$0.010	$&$-0.6	$&$1405	  $&$1642$&$[1567,1722]	$&$16.847	$\\
    $7	$&$0.001	$&$-0.9	$&$1405	  $&$1380$&$[1316,1449]	$&$1.808	$\\
    $7	$&$0.001	$&$-0.6	$&$1405	  $&$1400$&$[1358,1444]	$&$0.370	$\\ \hline
\end{tabular}
\caption{Transition times for small values of $\beta$, with various time
  steps $\Delta t$ and $z_{\rm min}$. Reference values are computed by DNS.}
\label{tab:meanoft1t2}
\end{table}

\begin{table}
  \centering
  \begin{tabular}{|c|c|c|c|c|} \hline
    $\beta	$&$\Delta t	$&$z_{\rm min}	$&Error on $\E(T3)$	&Error on
    $p$\\ \hline
    $1	$&$0.01	$&$-0.9	$&$1.73\%	$&$9.45\%$\\
$1	$&$0.01	$&$-0.8	$&$2.05\%	$&$5.81\%$\\
$1	$&$0.01	$&$-0.6	$&$1.76\%	$&$3.20\%$\\
$1	$&$0.001	$&$-0.9	$&$0.09\%	$&$2.84\%$\\
$1	$&$0.001	$&$-0.8	$&$0.19\%	$&$1.46\%$\\
$1	$&$0.001	$&$-0.6	$&$0.10\%	$&$0.17\%$\\
$5	$&$0.01	$&$-0.9	$&$0.97\%	$&$4.79\%$\\
$5	$&$0.01	$&$-0.6	$&$0.92\%	$&$4.23\%$\\
$5	$&$0.001	$&$-0.9	$&$0.39\%	$&$1.17\%$\\
$5	$&$0.001	$&$-0.6	$&$0.04\%	$&$1.35\%$\\
$7	$&$0.01	$&$-0.9	$&$0.54\%	$&$4.63\%$\\
$7	$&$0.01	$&$-0.6	$&$0.96\%	$&$5.97\%$\\
$7	$&$0.001	$&$-0.9	$&$0.21\%	$&$1.04\%$\\
$7	$&$0.001	$&$-0.6	$&$0.33\%	$&$1.17\%$\\ \hline
  \end{tabular}
  \caption{Comparison between DNS and the algorithm on the estimation of
    $\E(T_3)$ and of the probability $p$, for small values of $\beta$.}
  \label{tab:compa_transition_proba_cerou_dns}
\end{table}

\begin{table}
  \centering
  \begin{tabular}{|c|c|c|c|c|c|c|c|c|} \hline
$\beta	$&$p	$&$\E(T_1+T_3)	$&$\E(T_1+T_2)	$&$\E(T)
$&$\log(\E(T))/\beta$\\ \hline
$1	$&$1.350 \ 10^{-1}	$&$0.49930	$&$0.47300	$&$3.529	$&$1.261000$\\
$5	$&$1.240 \ 10^{-2}	$&$1.13658	$&$2.41300	$&$1.932\ 10^{2}	$&$1.052788$\\
$7	$&$3.347 \ 10^{-3}	$&$1.23931	$&$4.69608	$&$1.400\ 10^{3}	$&$1.034869$\\
$10	$&$1.411 \ 10^{-5}	$&$1.55896	$&$0.37247	$&$2.640\ 10^{4}	$&$1.018097$\\
$15	$&$1.239 \ 10^{-7}	$&$1.70723	$&$0.47593	$&$3.842\ 10^{6}	$&$1.010760$\\
$20	$&$1.007 \ 10^{-9}     $&$1.80644	$&$0.58504     $&$5.807\ 10^{8}	$&$1.008986$\\
$25	$&$8.381\ 10^{-12}	$&$1.89356	$&$0.70129     $&$8.367\ 10^{10}	$&$1.006007$\\
$30	$&$6.558\ 10^{-14}	$&$1.97645	$&$0.82802     $&$1.263\ 10^{13}	$&$1.005560$\\
$40	$&$4.043\ 10^{-18}	$&$2.11230	$&$1.12890     $&$2.792\ 10^{17}	$&$1.004270$\\ \hline
 \end{tabular}
  \caption{Computation of the mean transition time for various values of
    $\beta$. For small values of $\beta \leq 7 $, results come from Table~\ref{tab:meanoft1t2}, taking the best choice of $\Delta t$ and
    $z_{\rm min}$. For $\beta \geq 10$, the numerical parameters are
    $\Delta t= 0.001$, $N=10^5$ and $z_{\rm min} =-0.9$.}
  \label{tab:1d_res_proba_time}
\end{table}

\subsection{The two-dimensional case}
Let us go back to the two-dimensional case already used in Section~\ref{sec:2d}. Recall that the transition time is defined as
$$T=\inf \{t > 0, X_t \in B  \, \vert \, X_0 \sim \nu_{\partial A} \}.$$
Using the notation of Section~\ref{sec:1d_T}, let us also define
$$T_1=\inf \{t > 0, X_t \in  \Sigma_{z_{\rm min}} \, \vert \, X_0 \sim \nu_{\partial A}   \},$$
$$T_2=\inf \{t > 0, X_t  \in A  \, \vert \, X_0 \sim \nu_{\Sigma_{z_{\rm min}}} \text{ and
}\forall s \in (0,t), \  X_s \not \in  B \},$$
$$T_3=\inf \{t > 0, X_t \in B   \, \vert \, X_0 \sim \nu_{\Sigma_{z_{\rm min}}}\text{ and }
\forall s \in (0,t), \  X_s \not \in A \}.$$
We use again formula~\eqref{eq:esp_T_formula} to estimate $\E(T)$. The two
reaction coordinates $\xi_1$ and $\xi_2$ introduced in Section~\ref{sec:2d}
will be used. In all the computations, we take $z_{\rm min}=0.1$,
and $z_{\rm {max}}=0.9$ when the reaction coordinate is $\xi_1$ or $z_{\rm
  {max}}=1.5$ when the reaction coordinate is $\xi_2$.

First, in the case $\xi=\xi_2$ and $\beta = 1.67$, $p$ is estimated using the multilevel splitting
algorithm and a DNS. Results are presented in Table~\ref{tab:2d_res_p_beta167}, and show excellent agreement between the two estimates. We then check the consistency with the DNS on the transition
time estimation. The results are presented
in Table~\ref{tab:2d_res_T_beta167} for different values of the number of paths
$N$ and time-step size $\Delta t$. We again observe an excellent agreement.

\begin{figure}[htbp]
  \centering
  \epsfig{file=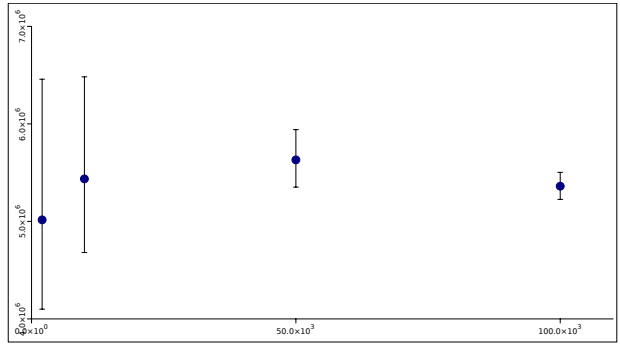,width=8cm}
  \caption{Convergence of the mean transition time $\E(T)$ as a function of $N$ for $\beta=6.67$,
    $\Delta t =0.01$ and  $\xi=\xi_2$.}
  \label{fig:convergence_2d_mean_time}
\end{figure}

Let us now consider the case  $\xi=\xi_2$ and $\beta=6.67$. Results are given in Table~\ref{tab:2d_res_beta667}. Note that in this case, $\E(T_3)$ is not significant in the estimation of
$\E(T)$. Figure~\ref{fig:convergence_2d_mean_time} illustrates the
convergence as $N$ increases. 

In order to compare with results obtained {\em via} the transition path theory in~\cite{metzner-schuette-vanden-eijnden-06}, we present in Table~\ref{tab:2d_res_rate_beta}  the estimation of
the reactive rate $k_{AB}$ from~$A$ to~$B$ which, by symmetry of the problem, is related to the mean transition time as: $k_{AB} = 2 / \E ( T )$. 
For	$\beta=1.67$, the rate computed {\em via} the multilevel splitting algorithm is consistent with those predicted from DNS and transition path
theory. For $\beta=6.67$ the rate is so small that any computation {\em via}
DNS would lead to totally unreasonable effort.
The estimate obtained using the multilevel splitting algorithm is very close from the
rate computed by transition path theory.

\begin{table}
  \centering
  \begin{tabular}{|c|c|c|c|c|} \hline
    $N$          &$\Delta t$&$p$              & C.I. on $p$               &$p$ \\ 
    $\times 10^3$&             &(algo)               & (algo)         &(DNS) \\ \hline
    $  2 $ &$0.01$  &$1.16  \ 10^{-2}$&$[1.04,1.28] \  10^{-2}$&$1.07  \ 10^{-2}$\\
    $10 $ &$0.01$  &$1.06  \ 10^{-2}$&$[1.05,1.07] \  10^{-2}$&$1.08  \ 10^{-2}$\\
    $50 $ &$0.01$  &$1.08  \ 10^{-2}$&$[1.08,1.08] \  10^{-2}$&$1.08  \ 10^{-2}$\\
    $100$&$0.01$  &$1.09  \ 10^{-2}$&$[1.09, 1.09] \  10^{-2}$&$1.08  \ 10^{-2}$\\
    $ 10 $&$0.001$&$5.42  \ 10^{-3}$&$[5.37, 5.47] \  10^{-3}$&$                    $\\
    $50  $&$0.001$&$5.43  \ 10^{-3}$&$[5.33, 5.54] \  10^{-3}$&$                    $\\ \hline
  \end{tabular}
  \caption{Computation of the probability $p$ of a reactive path for $\beta=1.67$ and $\xi=\xi_2$.}
  \label{tab:2d_res_p_beta167}
\end{table}

\begin{table}
  \centering
  \begin{tabular}{|c|c|c|c|c|c|c|c|} \hline
    $N$                  &$\Delta t$&$\E(T_1+T_2)$&$\E(T_1+T_3)$&$\E(T)$ &C.I. on & $\E(T)$ &Relative \\ 
    $\times 10^3$&             &(DNS) &(algo)&(algo) &$\E(T)$ &(DNS)  &error $(\%)$ \\ \hline
    $2    $  &$0.01$  &$2.72  \ 10^{-1}$&$1.43$&$24.66$&$[22.48;27.35]$&$26.54$&$7.61\%$\\
    $10  $  &$0.01$  &$2.73  \ 10^{-1}$&$1.56$&$27.14$&$[26.88;27.41]$&$26.50$&$2.38\%$\\
    $50  $  &$0.01$  &$2.74  \ 10^{-1}$&$1.54$&$26.66$&$[26.61;26.71]$&$26.53$&$0.48\%$\\
    $100$  &$0.01$  &$2.73  \ 10^{-1}$&$1.52$&$26.41$&$[26.38;26.44]$&$26.50$&$0.34\%$\\
    $10  $  &$0.001$&$1.42  \ 10^{-1}$&$1.32$&$27.32$&$[27.10;27.55]$&$26.49$&$3.05\%$\\
    $50  $  &$0.001$&$1.42  \ 10^{-1}$&$1.30$&$27.20$&$[26.73;27.70]$&$26.50$&$2.62\%$\\ \hline
\end{tabular}
\caption{Computation of the mean transition time for $\beta=1.67$ and  $\xi=\xi_2$. }
  \label{tab:2d_res_T_beta167}
\end{table}

\begin{table}
  \centering
  \begin{tabular}{|c|c|c|c|c|c|c|} \hline
    $N$&$\Delta t$&$p$&$\E(T_1+T_2)$&$\E(T_1+T_3)$&$\E(T)$ &C.I.  on \\ 
    $\times 10^3$  &                 &(algo)&(DNS)&(algo)&(algo) &$\E(T) $ \\ \hline
    $2$ &$0.01$ &$5.37 \ 10^{-8}$&$2.694  \ 10^{-1}$&$14.72$&$5.02 \ 10^{6}$&$[4.10;6.46]  \ 10^{6}$\\
    $10$      &$0.01$  &$4.96 \ 10^{-8}$&$2.694  \ 10^{-1}$&$15.94$&$5.44 \ 10^{6}$&$[4.68;6.48]  \ 10^{6}$\\
  $50$ &$0.01$  &$4.79 \ 10^{-8}$&$2.696  \ 10^{-1}$&$15.75$&$5.63 \ 10^{6}$& $[5.35;5.94]  \ 10^{6}$\\
    $100$     &$0.01$  &$5.03 \ 10^{-8}$&$2.698  \ 10^{-1}$&$14.71$&$5.36 \
    10^{6}$& $[5.22;5.50]  \ 10^{6}$\\
      $10$     &$0.001$&$4.50 \ 10^{-8}$&$2.036  \ 10^{-1}$&$14.67$&$4.52 \
    10^{6}$&$[3.78;5.62]  \ 10^{6}$\\ \hline
  \end{tabular}
  \caption{Computation of the mean transition time for $\beta=6.67$ and
    $\xi=\xi_2$.}
  \label{tab:2d_res_beta667}
\end{table}

\begin{table}
  \centering
  \begin{tabular}{|c|c|c|c|c|} \hline
    $N$&$\beta$&$\Delta t$&$k_{AB}$&C.I.  on \\ 
    $\times 10^3$  &    & &(algo) &$k_{AB}$  \\ \hline
    $ 2   	$&$1.67	$&$0.01	$&$2.03 \ 10^{-2} 	$&$[1.83;	2.22	] \ 10^{-2}$\\
    $10	$&$1.67	$&$0.01	$&$1.84 \ 10^{-2} 	$&$[1.82;	1.86	] \ 10^{-2} $\\
    $50	$&$1.67	$&$0.01	$&$1.88 \ 10^{-2} 	$&$[1.87;1.88] \ 10^{-2} $\\
    $100	$&$1.67	$&$0.01	$&$1.89 \ 10^{-2} 	$&$[1.89;1.90] \ 10^{-2} $\\
    $10	$&$1.67	$&$0.001$&$1.83 \ 10^{-2} 	$&$[1.81;1.84] \ 10^{-2} $\\
    $50	$&$1.67	$&$0.001$&$1.84 \ 10^{-2} 	$&$[1.80;1.87] \ 10^{-2} $\\ \hline
    $ 2	    $&$6.67	$&$0.01	$&$9.97 \ 10^{-8}	$&$[7.74;12.2] \ 10^{-8}$\\	
    $10	$&$6.67	$&$0.01	$&$9.20 \ 10^{-8}	$&$[7.71;10.7] \ 10^{-8}$\\
    $50	$&$6.67	$&$0.01	$&$8.88 \ 10^{-8}	$&$[8.42;9.34] \ 10^{-8}$\\	
    $100	$&$6.67	$&$0.01	$&$9.32 \ 10^{-8}	$&$[9.08;9.57] \ 10^{-8}$\\
    $10	$&$6.67	$&$0.001$&$1.11 \ 10^{-7}	$&$[8.89;13.2] \ 10^{-8}$\\ \hline
  \end{tabular}
  \caption{Reaction rate $\xi=\xi_2$. Values from~\cite{metzner-schuette-vanden-eijnden-06} are $k_{AB}=1.912 \ 10^{-2}$
    for $\beta = 1.67$ and $k_{AB}= 9.47 \ 10^{-8}$ for $\beta=6.67$.}
  \label{tab:2d_res_rate_beta}
\end{table}

Let us now consider the reaction coordinate $\xi_1(x,y)=x$. Results are given in Table~\ref{tab:2d_res_beta_xi1}. We observe that even though $\Sigma_{z_{\rm min}}$ is in this case very far from an isocommittor, the results are still consistent with those obtained above. Accordingly with the results of Section~\ref{sec:2d}, we note a larger variance on the results with this choice of the reaction coordinate, when $\beta=6.67$. The main source of variance is in the estimation $\hat{\alpha}_N$ of $p$.

\begin{table}
\centering
  \begin{tabular}{|c|c|c|c|c|c|c|c|} \hline
    $N$&$\beta$&$\Delta t$&$p$&$\E(T_1+T_2)$&$\E(T_1+T_3)$&$\E(T)$ &C.I.  on \\ 
    $\times 10^3$  &          &       &(algo)&(DNS)&(algo)&(algo) &$\E(T)$  \\ \hline
    $10	$&$1.67	$&$0.01	$&$2.15 \ 10^{-2}	$&$5.47 \ 10^{-1}$&$1.56	  $&$26.45           $&$[26.09;26.82]$\\
    $50	$&$1.67	$&$0.01	$&$2.14 \ 10^{-2}	$&$5.42 \ 10^{-1}$&$1.54	  $&$26.34           $&$[26.03;26.67]$\\
    $10	$&$6.67	$&$0.01	$&$2.61  \ 10^{-7}	$&$1.82	     	$&$11.22  $&$6.96 \
    10^{6}$&$[4.45;15.9 ] \ 10^{6}	$\\	
    $50  $&$6.67	$&$0.01	$&$2.92 \ 10^{-7}	$&$1.80	     	$&$11.24  $&$6.15 \
    10^{6}$&$[4.88;8.32] \ 10^{6} $\\ \hline
\end{tabular}
  \caption{Computation of the mean transition time with $\xi=\xi_1$.}
  \label{tab:2d_res_beta_xi1}
\end{table}

\section{Conclusion and perspectives}
\label{sec:conc}

We presented a multiple replica algorithm to generate an ensemble of reactive
trajectories. We illustrated its efficiency and accuracy on two test cases. We
also proposed an estimator of the transition times. Future works are of course required in order to test the interest of such an approach for larger systems.

In conclusion, we would like to mention two possible extensions of the approach. First, in a case where only the initial metastable state $A$ is known, one could think of using this algorithm to force the system to leave $A$ (without knowing the metastable states around {\em a priori}), by using as a reaction coordinate the distance to a reference configuration $x_A$ in $A$ (like $\xi_2$ above). The paths would then be generated until they go back to $A$, or they reach a given fixed final time $T$. This would generate equilibrium trajectories of time length~$T$, conditionally to reach a certain distance from $x_A$. It should be an efficient procedure to explore the energy landscape at a fixed positive temperature. Second, it would be interesting to test an adaptive procedure which, at the end of the algorithm, approximates the committor function thanks to the reactive trajectories, and then uses this approximation as a reaction coordinate, iteratively. In particular, note that  the isocommittors are also isolines of the function $\exp(\beta V) \rho$, where $\rho$ is the density along reactive paths which seems to be accurately obtained by our algorithm (see Section~\ref{sec:2d}). This should produce better and better results as the approximations of the isocommittors get more and more refined.

\medskip
{\bf Acknowledgements}: This work is supported by the Agence Nationale de la Recherche, under grants ANR-09-BLAN-0216-01 (MEGAS) and ANR-06-SETI-009 (Nebbiano), and by the ARC Hybrid. TL would like to thank Chris Chipot, Josselin Garnier and Rapha\"{e}l Roux for stimulating discussions on the subject.

\end{document}